\documentclass[12pt]{article}
\usepackage{amssymb,amsmath,epsfig}
\allowdisplaybreaks

\begin{document}
\title{\bf Anisotropic Spherical Solutions through Extended Gravitational
Decoupling Approach}
\author{M. Sharif \thanks{msharif.math@pu.edu.pk} and Qanitah Ama-Tul-Mughani
\thanks{qanitah94@gmail.com}\\
Department of Mathematics, University of the Punjab,\\
Quaid-e-Azam Campus, Lahore-54590, Pakistan.}

\date{}
\maketitle
\begin{abstract}
This paper is devoted to evaluating exact anisotropic spherical
solutions for static self-gravitating systems through extended
geometric deformation decoupling technique. For this purpose, we
consider an isotropic Tolman IV solution and extend it to
anisotropic domain by transforming both temporal as well as radial
metric potentials. To examine the physical viability and stability
of interior anisotropic solutions, we plot energy bounds, TOV
equation, causality condition and adiabatic index for the stars Her
X-I and PSR J 1416-2230. It is found that both obtained models show
realistic behavior as they fulfill all physical constraints as well
as stability criterion. We conclude that the extended gravitational
decoupling approach provides more proficient results to discuss the
interior configuration of stellar structures.
\end{abstract}

\section{Introduction}

General relativity as the geometric theory of gravitation provides
an elementary insight to the salient features of self-gravitating
objects. In astrophysics, the formulation of exact solutions of
Einstein field equations describe the interior distribution of
sellar structures. Schwarzschild \cite{1} found the spherical vacuum
solution which specifies the exterior region of perfect matter
distribution. Tolman \cite{2} computed several solutions for perfect
fluid in the presence of cosmological constant and investigated
smooth matching conditions of interior spacetime with the exterior
one. Lemaitre \cite{3} proposed that anisotropy in the interior of
celestial objects may arise due to phase transition, rotational
motion, presence of magnetic field or mixture of two fluids, etc. To
study the prominent features of anisotropic fluid distribution, many
physically acceptable solutions have been evaluated in relativistic
theories of gravity \cite{4}-\cite{4''''}.

The study of exact interior anisotropic solutions is a difficult
task due to non-linear characteristics of the evolution equations.
Over the years, several techniques have been proposed to obtain
physically acceptable models of celestial objects. In this context,
the gravitational decoupling through minimal geometric deformation
(MGD) approach has been used to formulate such analytical solutions
in cosmology as well as astrophysics. This approach was first
proposed by Ovalle \cite{9} to derive new consistent solutions for
static spherically symmetric spacetime in the background of
astrophysics braneworld. The MGD technique applies deformation on
the radial metric functions and splits the field equations in two
sets of differential equations that are convenient to solve as
compared to the original system. To attain the solution of the
complete model, the results of the decoupled equations are combined
using the principle of superposition. Following this procedure,
Ovalle and Linares \cite{10} computed an exact anisotropic solution
for compact spherical distribution and concluded that the model
corresponds to braneworld version of Tolman IV solution. Later,
Contreras and Bargue$\tilde{n}o$ \cite{17} applied this method on
1+2 static circularly symmetric spacetime and evaluated anisotropic
solutions from the static BTZ model.

Ovalle \cite{20} used the systematic approach of MGD to decouple
gravitational sources and obtained anisotropic spherical solutions
from the perfect fluid configuration. Ovalle et al. \cite{17a}
included the effects of anisotropy in corresponding isotropic
interior solution and derived three exact anisotropic models from
Tolman IV solution by means of MGD technique. Sharif and Sadiq
\cite{16} extended the singularity-free Krori-Barua solution to
anisotropic domain through this technique in the presence of the
electromagnetic field. In the same perspective, Gabbanelli et al.
\cite{16c} constructed new anisotropic solutions by taking
Durgapal-Fluoria isotropic fluid as an interior of stellar system.
Graterol \cite{16d} applied MGD decoupling phenomenon to generate
anisotropic analytic solutions from Buchdahl perfect fluid model for
static spherical self-gravitating system. Sharif and collaborators
\cite{16a,16b} obtained viable anisotropic solutions through this
procedure in modified gravity. Sharif and Ama-Tul-Mughani \cite{17'}
computed exact charged isotropic as well as anisotropic solutions in
a cloud of strings. Recently, Casadio et al. \cite{17''} used this
method to continuously isotropize the anisotropic solution with
vanishing complexity factor for the static sphere.

The MGD technique assists to study the essential characteristics of
stellar structures but it works under some limitations, e.g., the
geometric deformation can only be performed as long as the
interaction between the matter sources is purely gravitational.
Probably, its main drawback is that the transformation endures by
the metric component is minimal, i.e., it only modifies the radial
coordinate by leaving temporal metric potential as an invariant
quantity which may lead to certain shortcomings in the decoupling
phenomenon. To overcome this issue, Casadio et al. \cite{11}
introduced an extended version of MGD approach by employing the
deformation on both temporal as well as radial metric functions and
obtained a new solution for spherically symmetric spacetime.
However, this extension can only study the vacuum solutions as the
conservation law no longer holds in the presence of matter.
Consequently, the interior structure as well as intrinsic properties
of self-gravitating objects cannot be discussed through this
extended approach. Ovalle \cite{14} proposed a novel idea of
extended geometric deformation (EGD) by modifying both metric
potentials $(g_{tt}, g_{rr})$ which remains valid for entire
spacetime without depending upon the choice of matter distribution.
Contreras and Bargue$\tilde{n}$o \cite{14'} successfully decoupled
the field equations in 1+2-dimensional gravity through EGD technique
and implemented it to obtain exterior charged BTZ model from the
corresponding vacuum solution.

The aim of this paper is to study the most general extension of MGD
decoupling approach in the context of perfect spherical geometry. We
derive two exact anisotropic models from the well-known isotropic
Tolman IV solution by employing this extended decoupling phenomenon.
The paper is organized as follows. The next section provides the
basics of EGD approach and deals with the decoupling of the field
equations. In section \textbf{3}, we compute anisotropic solutions
by implying some physical constraints on the new gravitational
source. Section \textbf{4} is devoted to discussing the stability of
the resulting models and the final comments are summarized in the
last section.

\section{Gravitational Decoupled Field Equations}

The interior of spherically symmetric object in Schwarzschild
coordinates $(x^{\alpha})=(t,r,\theta,\phi)$ can be expressed as
\begin{equation}\label{1}
ds^{2}=-e^{\eta}dt^{2}+e^{\chi}dr^{2}+r^{2}(d\theta^2+\sin^{2}\theta
d\phi^2),
\end{equation}
where $\eta$ and $\chi$ have dependence on radial coordinate $r$
which varies from center to the boundary of star, i.e., $(0\leq
r\leq R)$. The energy-momentum tensor describing the internal
configuration of the stellar structure is taken as
\begin{equation}\label{2}
\tilde{T}_{\alpha}^{\beta}=T_{\alpha}^{\beta}+\sigma\vartheta_{\alpha}^{\beta},
\end{equation}
such that
\begin{equation}\label{2'}
T_{\alpha}^{\beta}=(\rho+p)u_{\alpha}u^{\beta}+p\delta_{\alpha}^{\beta},
\end{equation}
where $\rho$ stands for energy density, $p$ for pressure and
$u_{\alpha}$ corresponds to four velocity of the fluid. The factor
$\vartheta_{\alpha}^{\beta}$ describes an additional source which is
gravitationally coupled to the perfect fluid by a free parameter
$\sigma$. This source term generally produces anisotropy in the
stellar objects by incorporating new scalar, vector or tensor fields
in the respective model. The Einstein field equations for
non-generic gravitational sources explicitly read
\begin{eqnarray}\label{9}
T_{0}^{0}+\sigma\vartheta_{0}^{0}&=&-\frac{1}{r^2}-e^{-\chi}
\left(\frac{\chi^{'}}{r}-\frac{1}{r^2}\right),\\\label{10}
T_{1}^{1}+\sigma\vartheta_{1}^{1}&=&-\frac{1}{r^2}+e^{-\chi}
\left(\frac{\eta^{'}}{r}+\frac{1}{r^2}\right),\\\label{11}
T_{2}^{2}+\sigma\vartheta_{2}^{2}&=&\frac{e^{-\chi}}{4}
\left(2\eta^{''}+\eta^{'2}-\eta^{'}\chi^{'}
+\frac{2}{r}(\eta^{'}-\chi^{'})\right),
\end{eqnarray}
where prime means derivative with respect to $r$.

The conservation equation corresponding to metric (\ref{1}) turns
out to be
\begin{equation}\label{12}
p^{'}+\frac{\eta^{'}}{2}(\rho+p)+\sigma\left(\vartheta^{1'}_{1}
+\frac{\eta^{'}}{2}(\vartheta_{1}^{1}-\vartheta_{0}^{0})
+\frac{2}{r}(\vartheta_{1}^{1}-\vartheta_{2}^{2})\right)=0.
\end{equation}
Through direct analysis, the matter components can be defined as
\begin{eqnarray}\label{13}
\hat{\rho}=\rho-\sigma\vartheta_{0}^{0}, \quad
\hat{p}_{r}=p+\sigma\vartheta_{1}^{1}, \quad
\hat{p}_{t}=p+\sigma\vartheta_{2}^{2},
\end{eqnarray}
where $\hat{\rho}$ is the effective density, $\hat{p}_{r}$ and
$\hat{p}_{t}$ denote the effective radial and tangential pressures,
respectively. The anisotropy in the interior of celestial objects
takes the form
\begin{equation}\label{14}
\hat{\Delta}=\sigma(\vartheta_{2}^{2}-\vartheta_{1}^{1}),
\end{equation}
which indicates that the insertion of new source term
$\vartheta_{\alpha}^{\beta}$ produces anisotropy in self-gravitating
objects. Now, we have a system of three equations
(\ref{9})-(\ref{11}) with seven unknowns, i.e., two metric
components $\chi, \eta$ and five matter variables $\rho, p,
\vartheta_{0}^{0}, \vartheta_{1}^{1}, \vartheta_{2}^{2}$. To compute
the exact solution of the prescribed model, we use a novel approach
named as EGD and evaluate these unknown functions.

\subsection{Extended Geometric Decoupling Approach}

To solve the system of non-linear differential equations
(\ref{9})-(\ref{11}), we implement the EGD technique which
transforms the  field equations associated with source
$\vartheta_{\alpha}^{\beta}$ into an ``effective quasi-Einstein
system". We start by considering a known perfect fluid solution
$(\vartheta_{\alpha}^{\beta}=0)$ for the metric
\begin{eqnarray}\label{15}
ds^{2}=-e^{\xi(r)}dt^{2}+e^{\mu(r)}dr^{2}+r^{2}(d\theta^2
+\sin^{2}\theta d\phi^2),
\end{eqnarray}
where $e^{-\mu(r)}=1-\frac{2m(r)}{r}$ with the Misner-Sharp mass
$m(r)$. To analyze the impact of $\vartheta^{\alpha}_{\beta}$ on
$T^{\alpha}_{\beta}$, we consider the non-zero values of free
parameter and observe the following geometric deformations on the
metric functions, namely,
\begin{eqnarray}\label{16}
e^{-\mu}\rightarrow e^{-\chi}=e^{-\mu}+\sigma g(r), \quad
\xi\rightarrow \eta=\xi+\sigma h(r),
\end{eqnarray}
where $g(r)$ and $h(r)$ denote the deformation functions subject to
the radial and temporal coordinates, respectively. We substitute the
above decompositions in Eqs.(\ref{9})-(\ref{11}) to split them into
two set of differential equations. The first set contains the
standard field equations for $T^{\alpha}_{\beta}$ given by
\begin{eqnarray}\label{17}
\rho&=&\frac{1}{r^2}+e^{-\mu}\left(\frac{\mu^{'}}{r}-\frac{1}
{r^2}\right),\\\label{18} p&=&-\frac{1}{r^2}+e^{-\mu}
\left(\frac{\xi^{'}}{r}+\frac{1}{r^2}\right),\\\label{19}
p&=&e^{-\mu}\left(\frac{\xi^{''}}{2}+\frac{\xi^{'2}}{4}
-\frac{\xi^{'}\mu^{'}}{4}+\frac{1}{2r}(\xi^{'}-\mu^{'})\right),
\end{eqnarray}
while the second set consists of evolution equations for
$\vartheta_{\alpha}^{\beta}$
\begin{eqnarray}\label{20}
\vartheta_{0}^{0}&=&\frac{g}{r^2}+\frac{g^{'}}{r},\\ \label{21}
\vartheta_{1}^{1}&=&
g\left(\frac{1}{r^2}+\frac{\eta^{'}}{r}\right)+\frac{e^{-\mu}h^{'}}{r},\\
\nonumber \vartheta_{2}^{2}&=&\frac{g}{4}\left(2\eta^{''}+\eta^{'2}
+\frac{2\eta^{'}}{r}\right)+\frac{g^{'}}{4}\left(\eta^{'}+\frac{2}{r}\right)
+\frac{e^{-\mu}}{4}\left(2h^{''}+\sigma h^{'2}
+\frac{2h^{'}}{r}\right.\\\label{22}&+&\left.2\xi^{'}h^{'}-\mu^{'}h^{'}\right).
\end{eqnarray}
The continuity equation, $(\tilde{T}_{\alpha}^{\beta})_{;\beta}=0$,
for the line-element (\ref{15}) yields
\begin{equation}\label{23}
p^{'}+\frac{\xi^{'}}{2}(\rho+p)+\frac{\sigma
h^{'}}{2}(\rho+p)+\sigma
\left(\vartheta_{1}^{1'}+\frac{\eta^{'}}{2}(\vartheta_{1}^{1}
-\vartheta_{0}^{0})+\frac{2}{r}(\vartheta_{1}^{1}-\vartheta_{2}^{2})\right)=0.
\end{equation}
Setting $h^{'}(r)=0$ in the above identity leads to the specific
case of MGD technique and the system of equations
(\ref{20})-(\ref{22}) reduce to the quasi-Einstein system \cite{20,
21}.

Now, we study the critical behavior of conservation equation to
obtain a successful decoupling approach. Since the Bianchi identity
for the perfect fluid configuration remains preserved for the metric
$(\xi,\mu)$, so that
\begin{eqnarray}\label{24}
\nabla_{\alpha}^{(\xi,\mu)}T_{\beta}^{\alpha}=0,
\end{eqnarray}
whereas the conservation of $T_{\beta}^{\alpha}$ corresponding to
metric (\ref{1}) takes the form
\begin{eqnarray}\label{25}
\nabla_{\alpha}T_{\beta}^{\alpha}=
\nabla_{\alpha}^{(\xi,\mu)}T_{\beta}^{\alpha}+\frac{\sigma
h^{'}}{2}(\rho+p)\delta_{\beta}^{1}.
\end{eqnarray}
Using Eq.(\ref{24}) in the above expression gives
\begin{eqnarray}\label{26}
\nabla_{\alpha}T_{\beta}^{\alpha}=\frac{\sigma
h^{'}}{2}(\rho+p)\delta_{\beta}^{1},
\end{eqnarray}
while the divergence of the gravitational source
$\vartheta^{\alpha}_{\beta}$ reads
\begin{eqnarray}\label{27}
\nabla_{\alpha}\vartheta_{\beta}^{\alpha}=-\frac{
h^{'}}{2}(\rho+p)\delta_{\beta}^{1},
\end{eqnarray}
which can be expressed as the linear combination of
Eqs.(\ref{20})-(\ref{22}), so we have
\begin{eqnarray}\label{28}
\vartheta_{1}^{1'}+\frac{\eta^{'}}{2}(\vartheta_{1}^{1}-\vartheta_{0}^{0})
+\frac{2}{r}(\vartheta_{1}^{1}-\vartheta_{2}^{2})=-\frac{
h^{'}}{2}(\rho+p).
\end{eqnarray}
From the above identities (\ref{26}) and (\ref{27}), we can conclude
that matter sources $T_{\beta}^{\alpha}$ and
$\vartheta_{\beta}^{\alpha}$ can be decoupled successfully until the
energy is able to transform from one source to another. However, the
MGD approach allows a purely gravitational interaction between
matter sources by restricting the exchange of energy between them.
It is worthwhile to mention here that EGD approach can also work
without exchange of energy in merely two scenarios: either the fluid
is barotropic, i.e., $T_{0}^{0}=T_{1}^{1}$ or the isotropic sector
represents the vacuum solution, i.e., $T_{\alpha}^{\beta}=0$.

The equations of motion for anisotropic source (\ref{20})-(\ref{22})
can be identified as the field equations for anisotropic conserved
tensor $\vartheta^{*\alpha}_{\beta}$ defined by
\begin{eqnarray}\label{29}
\vartheta^{*\alpha}_{\beta}=\vartheta^{\alpha}_{\beta}
-\frac{1}{r^2}\delta_{\beta}^{0}\delta^{\alpha}_{0}
-\left(X_{1}+\frac{1}{r^2}\right)\delta_{\beta}^{1}\delta^{\alpha}_{1}
-X_{2}\delta_{\beta}^{2}\delta^{\alpha}_{2},
\end{eqnarray}
which is explicitly written as
\begin{eqnarray}\label{30}
\vartheta^{*0}_{0}&=&\vartheta^{0}_{0}-\frac{1}{r^2},\\ \label{30'}
\vartheta^{*1}_{1}&=&\vartheta^{1}_{1}-\left(X_{1}+\frac{1}{r^2}
\right),\\\label{30''} \vartheta^{*2}_{2}&=&\vartheta^{2}_{2}-X_{2},
\end{eqnarray}
with
\begin{eqnarray}\nonumber
X_{1}&=&\frac{e^{-\mu}h^{'}}{r}, \quad
X_{2}=\frac{e^{-\mu}}{4}\left(2h^{''}+\sigma
h^{'2}+\frac{2h^{'}}{r}+2\xi^{'}h^{'}-\mu^{'}h^{'}\right).
\end{eqnarray}
In this scenario, the continuity equation turns out to be
\begin{eqnarray}\label{31}
(\vartheta^{*1}_{1})^{'}+\frac{\eta^{'}}{2}\left(\vartheta^{*1}_{1}
-\vartheta^{*0}_{0}\right)+
\frac{2}{r}\left(\vartheta^{*1}_{1}-\vartheta^{*2}_{2}\right)=0,
\end{eqnarray}
with the metric
\begin{eqnarray}\label{32}
ds^{2}=-e^{\eta}dt^{2}+\frac{dr^{2}}{g(r)}+r^{2}(d\theta^2+\sin^{2}\theta
d\phi^2).
\end{eqnarray}

\section{Anisotropic Solutions}

To derive the anisotropic solutions for the interior of
self-gravitating system, we solve Eqs.(\ref{20})-(\ref{22}) through
EGD approach. In this regard, we first turn off the free parameter
and consider a known solution for perfect matter distribution in the
background of spherically symmetric spacetime. We choose a
well-known Tolman IV model for perfect fluid given by \cite{23a}
\begin{eqnarray}\label{39}
e^{\xi}&=&\mathcal{B}^2(1+\frac{r^2}{\mathcal{A}^2}),\\
\label{40} e^{\mu}&=&\frac{1+\frac{2r^2}{\mathcal{A}^2}}
{(1-\frac{r^2}{\mathcal{C}^2})(1+\frac{r^2}{\mathcal{A}^2})},\\
\label{41}
\rho&=&\frac{3\mathcal{A}^4+\mathcal{A}^2(3\mathcal{C}^2+7r^2)
+2r^2(\mathcal{C}^2+3r^2)}{\mathcal{C}^2(\mathcal{A}^2+2r^2)^2},\\
\label{42} p&=&\frac{\mathcal{C}^2-\mathcal{A}^2-3r^2}{\mathcal{C}^2
(\mathcal{A}^2+2r^2)}.
\end{eqnarray}
The integration constants $\mathcal{A}$, $\mathcal{B}$ and
$\mathcal{C}$ are evaluated by the smooth matching of interior and
exterior spacetimes. Here, we choose Schwarzschild as an exterior
metric which yields
\begin{eqnarray}\label{44}
\mathcal{A}^2&=&\frac{R^2(R-3\mathcal{M})}{\mathcal{M}}, \quad
\mathcal{B}^2=1-\frac{3\mathcal{M}}{R}, \quad
\mathcal{C}^2=\frac{R^3}{\mathcal{M}},
\end{eqnarray}
with the compactness factor $\frac{\mathcal{M}}{2R}=\frac{2}{9}$.
The above expressions ensure the continuity of isotropic solution
with the exterior geometry at the boundary $(r=R)$. However, the
inclusion of gravitational source $\vartheta_{\alpha}^{\beta}$ in
the perfect fluid distribution will modify the respective
quantities, accordingly.

To obtain the anisotropic solutions, we take $\sigma\neq 0$ in the
interior of spherical object and solve the equations of motion
(\ref{20})-(\ref{22}). These equations interlink the five unknown
functions, namely, geometric deformations $(h(r), g(r))$ and source
term $\vartheta_{\alpha}^{\beta}$. Hence, we need some additional
constraints to minimize the number of unknowns. For this purpose, we
impose an equation of state and two physical constraints on the
components of $\vartheta_{\alpha}^{\beta}$ to obtain exact as well
as viable models in the following subsections.

\subsection{Solution I}

In order to compute exact anisotropic solution, we implement a
linear equation of state on $\vartheta_{\alpha}^{\beta}$ as
\begin{equation}\label{47}
\vartheta_{0}^{0}=\delta\vartheta_{1}^{1}+\gamma\vartheta_{2}^{2},
\end{equation}
and apply a constraint on $\vartheta_{1}^{1}$ to close the system.
From the junction conditions of Schwarzschild spacetime with the
interior geometry, we get $p(R)\sim -\sigma
(\vartheta_{1}^{1}(R))_{-}$. Thus the suitable choice is taken to be
\begin{equation}\label{48}
\vartheta_{1}^{1}=-p.
\end{equation}
Here, we set $\delta=1$ and $\gamma=0$ which lead to the relation
$\vartheta_{0}^{0}=\vartheta_{1}^{1}$. Using the field
Eqs.(\ref{20}) and (\ref{21}), the deformation functions are
evaluated as
\begin{eqnarray}\label{49}
g&=&\frac{-2r\left(\mathcal{A}^2+2\mathcal{C}^2-2
r^2\right)+\sqrt{2}\mathcal{A}\left(\mathcal{A}^2+2\mathcal{C}^2\right)
\tan^{-1}\left(\frac{\sqrt{2}
r}{\mathcal{A}}\right)}{8\mathcal{C}^2r},\\\nonumber
h&=&\int\frac{1}{\varepsilon}\left(2 r
\left(\mathcal{A}^6+8\mathcal{C}^2r^4+6
\mathcal{A}^2r^2\left(\mathcal{C}^2+2r^2\right)+\mathcal{A}^4
\left(2\mathcal{C}^2+7r^2\right)\right)-\sqrt{2}\mathcal{A} \right.\\
\label{50}&\times&\left.\left(\mathcal{A}^2+2\mathcal{C}^2\right)
\left(\mathcal{A}^4+5\mathcal{A}^2
r^2+6r^4\right)\tan^{-1}\left(\frac{\sqrt{2}
r}{\mathcal{A}}\right)\right)dr,
\end{eqnarray}
where
\begin{eqnarray}\nonumber
\varepsilon&=&r\left(\mathcal{A}^2+r^2\right)
(8(\mathcal{C}-r)r(\mathcal{C}+r)
\left(\mathcal{A}^2+r^2\right)+\left(\mathcal{A}^2+2r^2\right)
\sigma(4r^3\\\nonumber&-&2\left(\mathcal{A}^2+2\mathcal{C}^2
\right)r+\sqrt{2}\mathcal{A}
\left(\mathcal{A}^2+2\mathcal{C}^2\right)
\tan^{-1}\left(\frac{\sqrt{2}r}{\mathcal{A}}\right))),
\end{eqnarray}
which yield the temporal as well as radial metric components as
\begin{eqnarray}\nonumber
\eta&=&\ln\left(\mathcal{B}^2(1+\frac{r^2}{\mathcal{A}^2})
\right)+\sigma\int\frac{1}{\varepsilon}\left(2r
\left(\mathcal{A}^6+8\mathcal{C}^2r^4+6
\mathcal{A}^2r^2\left(\mathcal{C}^2+2r^2\right)\right.\right.\\
\nonumber&+&\left.\left.\mathcal{A}^4
\left(2\mathcal{C}^2+7r^2\right)\right)-\sqrt{2}\mathcal{A}
\left(\mathcal{A}^2+2\mathcal{C}^2\right)\left(\mathcal{A}^4
+5\mathcal{A}^2r^2+6r^4\right)\right.\\
\label{51}&\times&\left.\tan^{-1}\left(\frac{\sqrt{2}
r}{\mathcal{A}}\right)\right)dr,\\\nonumber
e^{-\chi}&=&\frac{(1-\frac{r^2}{\mathcal{C}^2})(1+\frac{r^2}
{\mathcal{A}^2})}{1+\frac{2r^2}{\mathcal{A}^2}}+\frac{\sigma}
{8\mathcal{C}^2r}(\sqrt{2}\mathcal{A}\left(\mathcal{A}^2
+2\mathcal{C}^2\right)\tan^{-1}\left(\frac{\sqrt{2}
r}{\mathcal{A}}\right)\\ \label{52}
&-&2r\left(\mathcal{A}^2+2\mathcal{C}^2-2 r^2\right)).
\end{eqnarray}
Notice that for $\sigma=0$, the above equations reduce to the
standard Tolman IV spherical solution for perfect matter
distribution.

To measure the effects of anisotropy on the constants
$(\mathcal{A},\mathcal{B},\mathcal{C})$, we evaluate their
expressions through matching conditions. The continuity of the first
fundamental form leads to the following identities
\begin{eqnarray}\nonumber
\ln\left(1-\frac{2\mathcal{M}}{R}\right)&=&
\ln\left(\mathcal{B}^2(1+\frac{R^2}{\mathcal{A}^2})\right)+\sigma\bigg(
\int\frac{1}{\varepsilon}\left(2 r
\left(\mathcal{A}^6+8\mathcal{C}^2r^4+6
\mathcal{A}^2r^2\right.\right.\\
\nonumber&\times&\left.\left.\left(\mathcal{C}^2+2r^2\right)
+\mathcal{A}^4\left(2\mathcal{C}^2+7r^2\right)\right)-\sqrt{2}
\mathcal{A}\left(\mathcal{A}^2+2\mathcal{C}^2\right)\right.\\
\label{53}&\times&\left.\left(\mathcal{A}^4+5\mathcal{A}^2
r^2+6r^4\right)\tan^{-1}\left(\frac{\sqrt{2}
r}{\mathcal{A}}\right)\right)\bigg)_{r=R}dr,\\
\nonumber1-\frac{2\mathcal{M}}{R}&=&\frac{(1-\frac{R^2}{\mathcal{C}^2})
(1+\frac{R^2}{\mathcal{A}^2})}
{1+\frac{2R^2}{\mathcal{A}^2}}+\frac{\sigma}{8\mathcal{C}^2R}
(\sqrt{2}\mathcal{A}\left(\mathcal{A}^2+2\mathcal{C}^2\right)
\tan^{-1}\left(\frac{\sqrt{2} R}{\mathcal{A}}\right)\\ \label{54}
&-&2R\left(\mathcal{A}^2+2\mathcal{C}^2-2 R^2\right)),
\end{eqnarray}
whereas the continuity of second fundamental form
$(p(R)+\sigma(\vartheta_{1}^{1}(R))_{-}=0)$ yields
\begin{equation}\label{55}
p(R)=0 \quad\Rightarrow\quad \mathcal{C}^2=3R^2+\mathcal{A}^2.
\end{equation}
These equations (\ref{53})-(\ref{55}) are the necessary and
sufficient conditions for the smooth matching of exterior and
interior spacetimes. In the case of pressure-like constraint, the
matter variables are evaluated as
\begin{eqnarray}\label{56}
\hat{\rho}&=&\frac{3\mathcal{A}^4+\mathcal{A}^2(3\mathcal{C}^2+7r^2)
+2r^2(\mathcal{C}^2+3r^2)}{\mathcal{C}^2(\mathcal{A}^2+2r^2)^2}
+\sigma\left(\frac{\mathcal{C}^2-\mathcal{A}^2-3r^2}
{\mathcal{C}^2(\mathcal{A}^2+2r^2)}\right),\\\label{57}
\hat{p}_{r}&=&\frac{\mathcal{C}^2-\mathcal{A}^2-3r^2}
{\mathcal{C}^2(\mathcal{A}^2+2r^2)}(1 - \sigma),\\
\nonumber\hat{p}_{t}&=&\frac{\mathcal{C}^2-\mathcal{A}^2-3r^2}
{\mathcal{C}^2(\mathcal{A}^2+2r^2)}+\sigma
(2r(16r^4(\mathcal{C}^4-3\mathcal{C}^2r^2
(-1+\sigma)+3r^4(-1+\sigma))\\
\nonumber&+&\mathcal{A}^8(1-2\sigma)+4\mathcal{A}^4
(\mathcal{C}^2r^2(21-8\sigma)+\mathcal{C}^4(-1+\sigma)
-r^4(11+\sigma))\\
\nonumber&+&4
\mathcal{A}^2r^2(2\mathcal{C}^2r^2(17-10\sigma)+12r^4(-2+\sigma
)+\mathcal{C}^4(1+2\sigma))\\
\nonumber&-&\mathcal{A}^6(2\mathcal{C}^2
(-6+\sigma)+r^2(1+12\sigma)))+\sqrt{2} \mathcal{A}
(\mathcal{A}^2+2\mathcal{C}^2)(\mathcal{A}^2+2 r^2)\\
\nonumber&\times&(-(\mathcal{A}^2+2\mathcal{C}^2)
(\mathcal{A}^2+3r^2)+2(\mathcal{A}^4+6r^4-\mathcal{A}^2
(\mathcal{C}^2-6r^2))\sigma)\tan^{-1}(\frac{\sqrt{2}
r}{\mathcal{A}}))\\ \nonumber&\times&\left(2
\mathcal{C}^2\left(\mathcal{A}^2+2r^2\right)^2[8(\mathcal{C}-r)r
(\mathcal{C}+r)
\left(\mathcal{A}^2+r^2\right)+\left(\mathcal{A}^2+2r^2\right)
\right.\\ \label{58}&\times&\left.\sigma
(-2\left(\mathcal{A}^2+2\mathcal{C}^2\right)r+4
r^3+\sqrt{2}\mathcal{A}\left(\mathcal{A}^2+2\mathcal{C}^2\right)
\tan^{-1}\left(\frac{\sqrt{2}
r}{\mathcal{A}}\right))]\right)^{-1},\\ \nonumber
\hat{\Delta}&=&(\mathcal{A}^2+2
\mathcal{C}^2)\sigma(2r(\mathcal{A}^6+2\mathcal{A}^2r^2
\left(2r^2+\mathcal{C}^2(7-2\sigma)\right)\\ \nonumber
&+&\mathcal{A}^4(2 \mathcal{C}^2+r^2(7-2
\sigma))-8r^4(\mathcal{C}^2(-2+\sigma)-r^2(-1+\sigma )))\\
\nonumber&-&\sqrt{2}\mathcal{A}\left(\mathcal{A}^2+2
\mathcal{C}^2\right) \left(\mathcal{A}^2+2 r^2\right)
\left(\mathcal{A}^2+r^2(3-2\sigma
)\right)\tan^{-1}\left(\frac{\sqrt{2} r}{\mathcal{A}}\right))\\
\nonumber&\times&(2\mathcal{C}^2(\mathcal{A}^2+2 r^2)^2(8
(\mathcal{C}-r) r (\mathcal{C}+r)
(\mathcal{A}^2+r^2)+(\mathcal{A}^2+2 r^2)\\ \label{59}&\times&
\sigma\left(-2\left(\mathcal{A}^2+2\mathcal{C}^2\right) r+4
r^3+\sqrt{2}\mathcal{A} \left(\mathcal{A}^2+2\mathcal{C}^2\right)
\tan^{-1}\left(\frac{\sqrt{2} r}{\mathcal{A}}\right)\right)))^{-1}.
\end{eqnarray}
The corresponding mass function takes the form
\begin{eqnarray*}
m&=&4\pi \int r^2\hat{\rho} dr, \\&=&\frac{\pi  (\sigma -1)
\left(\sqrt{2} \mathcal{A} \left(\mathcal{A}^2+2
\mathcal{C}^2\right) \tan ^{-1}\left(\frac{\sqrt{2}
r}{\mathcal{A}}\right)-2 r \left(\mathcal{A}^2+2 \mathcal{C}^2-2
r^2\right)\right)}{2 \mathcal{C}^2},
\end{eqnarray*}
while the central density and pressure are given by
\begin{eqnarray*}
\hat{\rho}_{c}&=&\frac{(1-\sigma ) \left(3 \mathcal{A}^4+3
\mathcal{A}^2 \mathcal{C}^2\right)}{\mathcal{A}^4 \mathcal{C}^2}, \\
\hat{p}_{rc}&=&\frac{(1-\sigma)
\left(\mathcal{C}^2-\mathcal{A}^2\right)}{\mathcal{A}^2
\mathcal{C}^2},
\end{eqnarray*}
which have positive as well as finite characteristics for
$\sigma<1$.
\begin{figure}\center
\epsfig{file=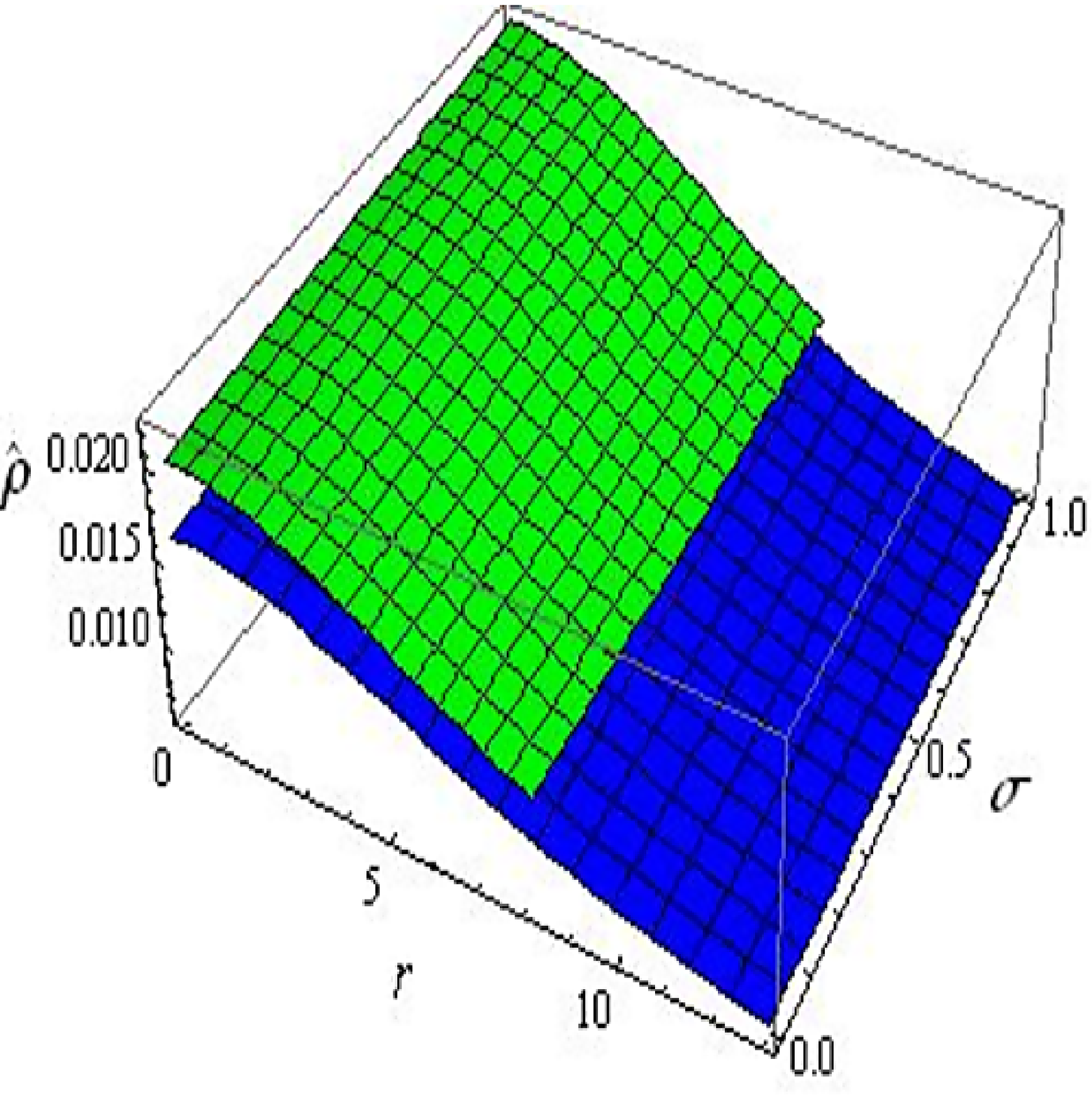,width=0.45\linewidth}\epsfig{file=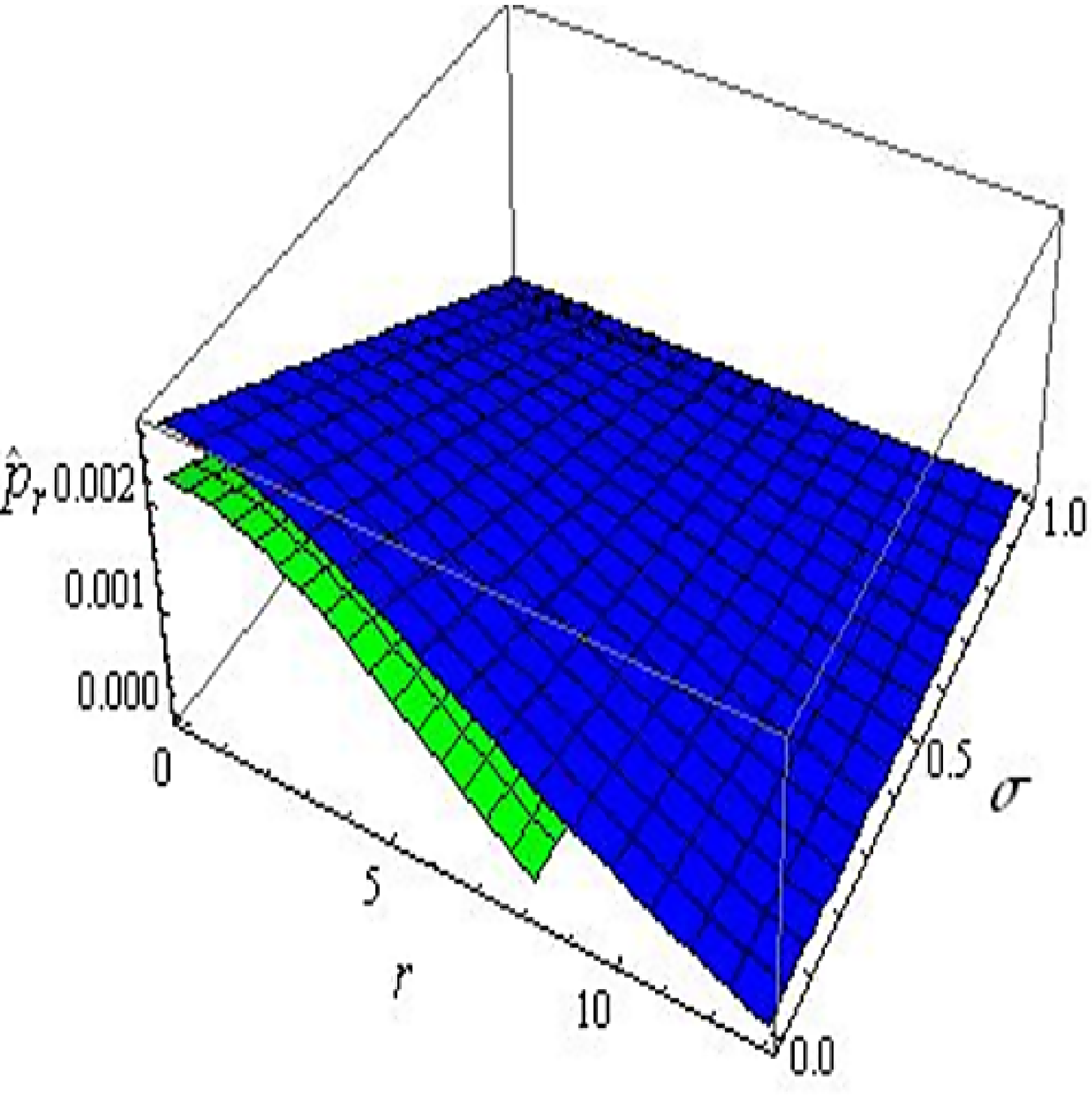,
width=0.45\linewidth} \epsfig{file=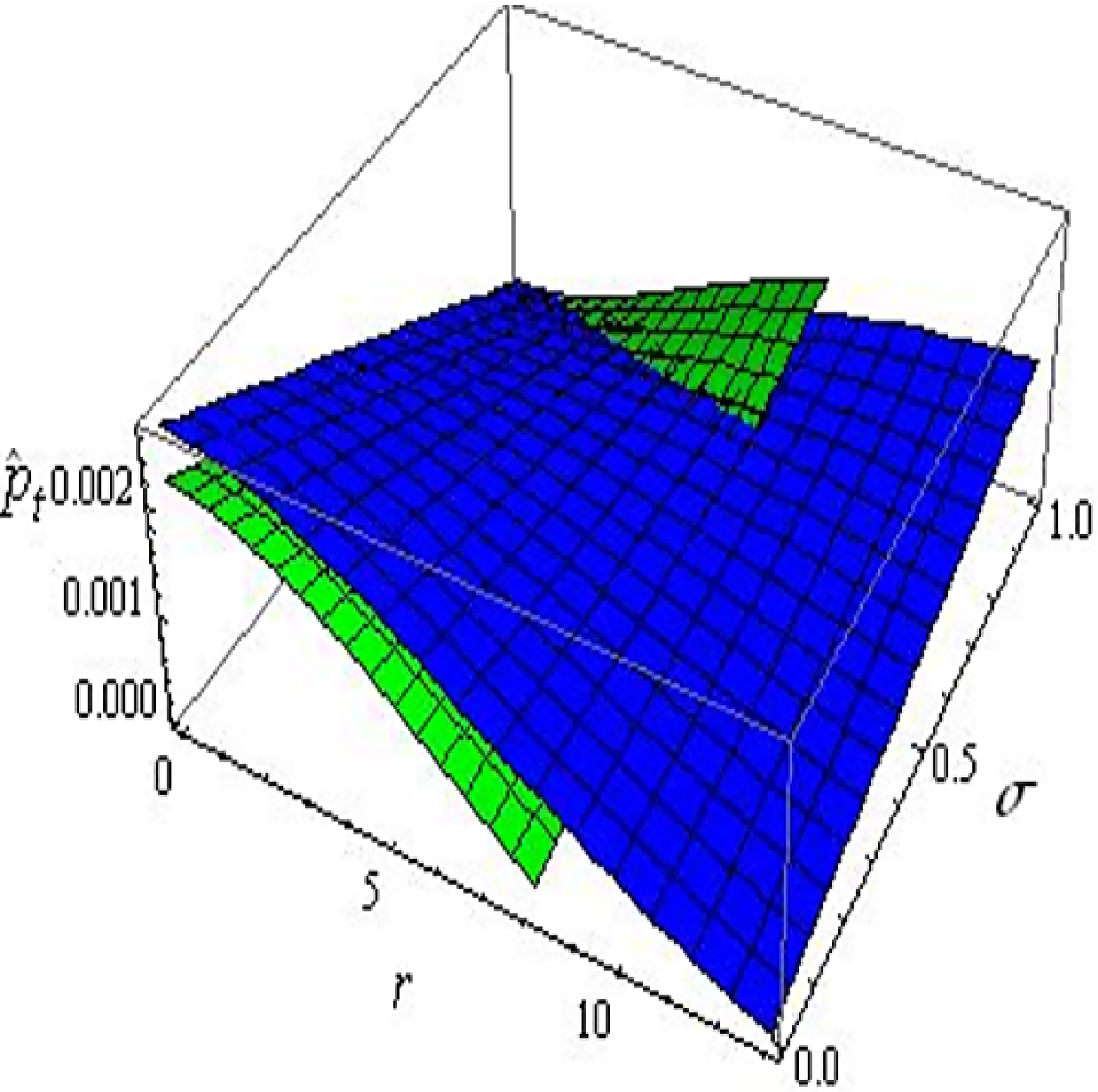,width=0.45\linewidth}
\epsfig{file=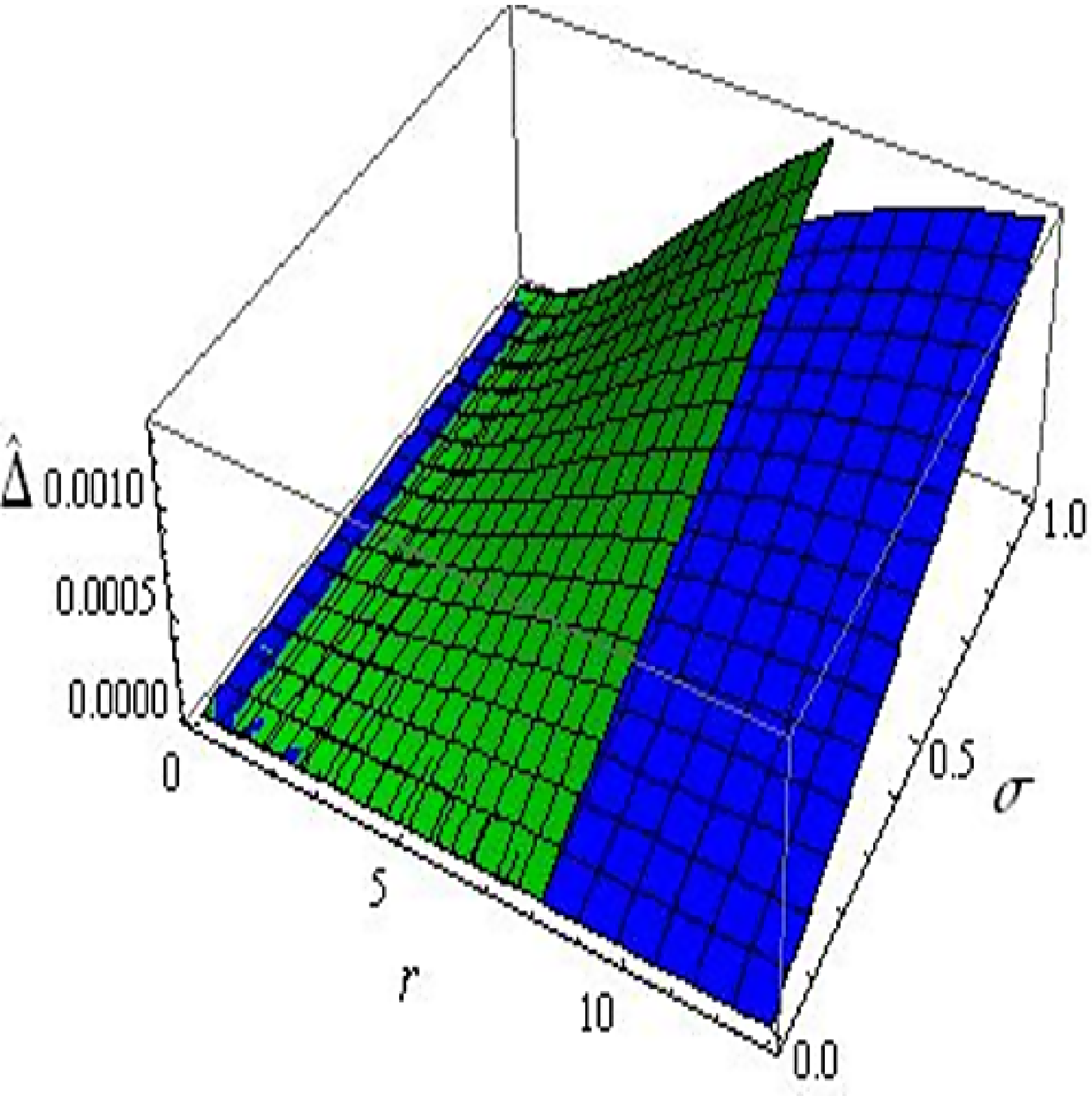,width=0.45\linewidth} \caption{Plots of
$\hat{\rho}$, $\hat{p}_{r}$, $\hat{p}_{t}$ and $\hat{\Delta}$ for
Her X-I (green) and PSR J 1416-2230 (blue) for solution
$\mathbf{I}$.}
\end{figure}
\begin{figure}\center
\epsfig{file=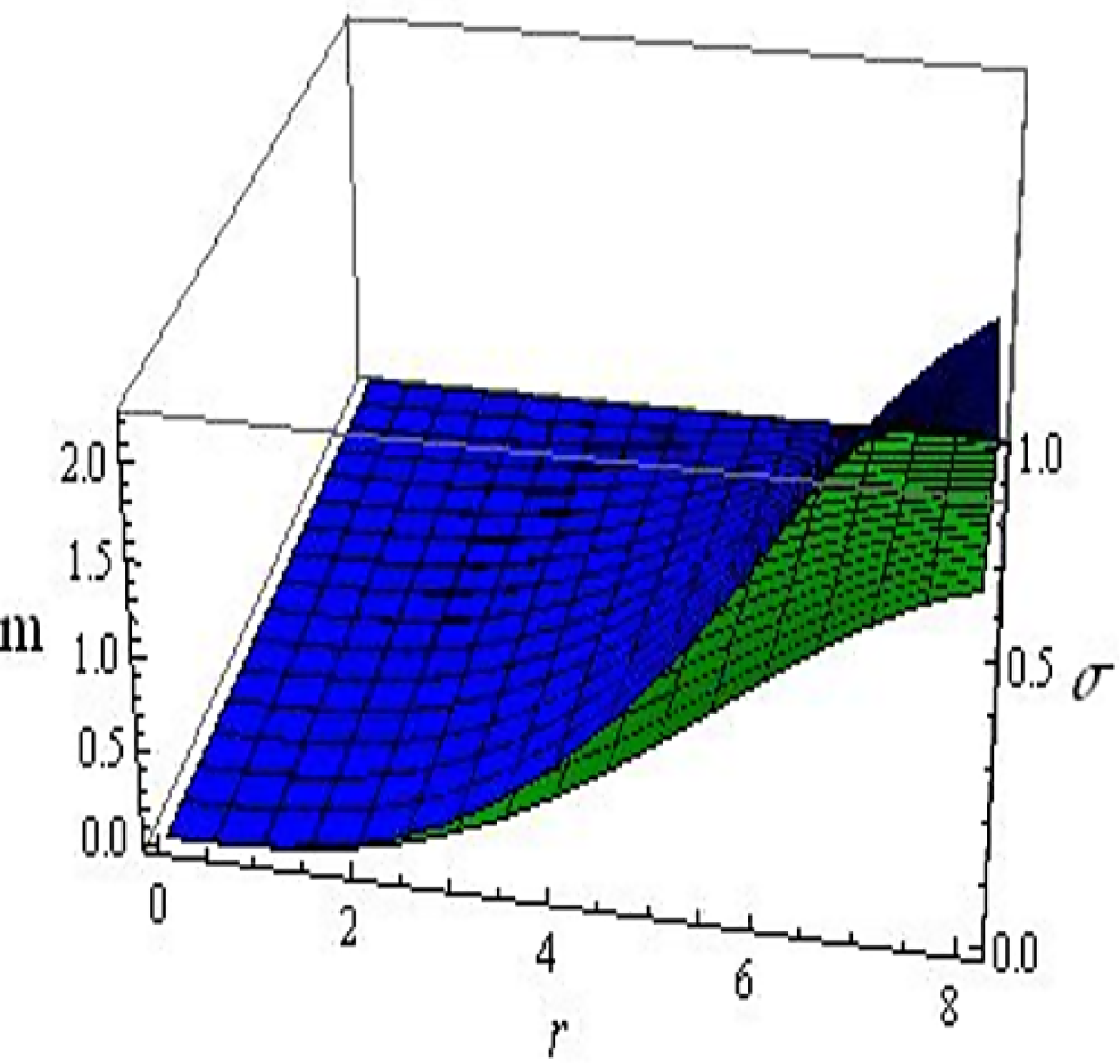,width=0.45\linewidth} \caption{Plot of mass for
Her X-I (green) and PSR J 1416-2230 (blue) for solution
$\mathbf{I}$.}
\end{figure}

We plot the above solution to analyze the physical characteristic of
matter variables for the stars Her X-I ($\mathcal{M}=1.25375
\text{km}$, $R=8.10 \text{km}$) and PSR J 1416-2230
($\mathcal{M}=2.90575 \text{km}$, $R=13 \text{km}$). Here, we choose
$\mathcal{A}$ as a free parameter whose value is presented in
isotropic solution (\ref{44}) while the expression of $\mathcal{C}$
is taken from Eq.(\ref{55}). For the feasible behavior of stellar
structure, the energy density and pressures $(\hat{p}_{r},
\hat{p}_{t})$ must be positive, finite as well as maximum at the
center of star. The profiles of energy density and anisotropic
pressures are displayed in Figure \textbf{1} which indicate that
density has the maximum value in the interior of compact star
$(r=0)$ with monotonically decreasing behavior towards the surface.
The sketch of $\hat{p}_{r}$ and $\hat{p}_{t}$ observe the same
physical trend as that of energy density and ultimately become zero
at the boundary of the star. From the plot of anisotropic parameter
(right plot, second row of Figure \textbf{1}), one can easily
investigate that the anisotropy vanishes at the center of
self-gravitating object and attains a maximum towards its surface.
Moreover, it is noted that the decoupling parameter $\sigma$
decreases the ranges of anisotropic pressures $(\hat{p}_{r},
\hat{p}_{t})$ whereas the density as well as anisotropy in the
system slightly increases under its effects. Figure \textbf{2}
indicates that the mass of stellar structure has direct relation
with its radius whereas the decoupling parameter depicts the inverse
scenario.

To measure the viability of developed solution, we examine the
physical behavior of energy bounds. These bounds are some
restrictions that are enforced on the energy-momentum tensor to
identify the realistic fluid distribution. For anisotropic
configuration, these conditions turn out to be
\begin{eqnarray}\nonumber
\hat{\rho}\geq 0, \quad \hat{\rho}+\hat{p}_{r}\geq0,\\ \nonumber
\hat{\rho}+\hat{p}_{t}\geq0, \quad \hat{\rho}-\hat{p}_{r}\geq0,\\
\nonumber \hat{\rho}-\hat{p}_{t}\geq0, \quad
\hat{\rho}+\hat{p}_{r}+2\hat{p}_{t}\geq0.
\end{eqnarray}
It is found that anisotropic model satisfies all energy conditions
which ensure the viability of our constructed solution as shown in
Figure \textbf{3}.
\begin{figure}\center
\epsfig{file=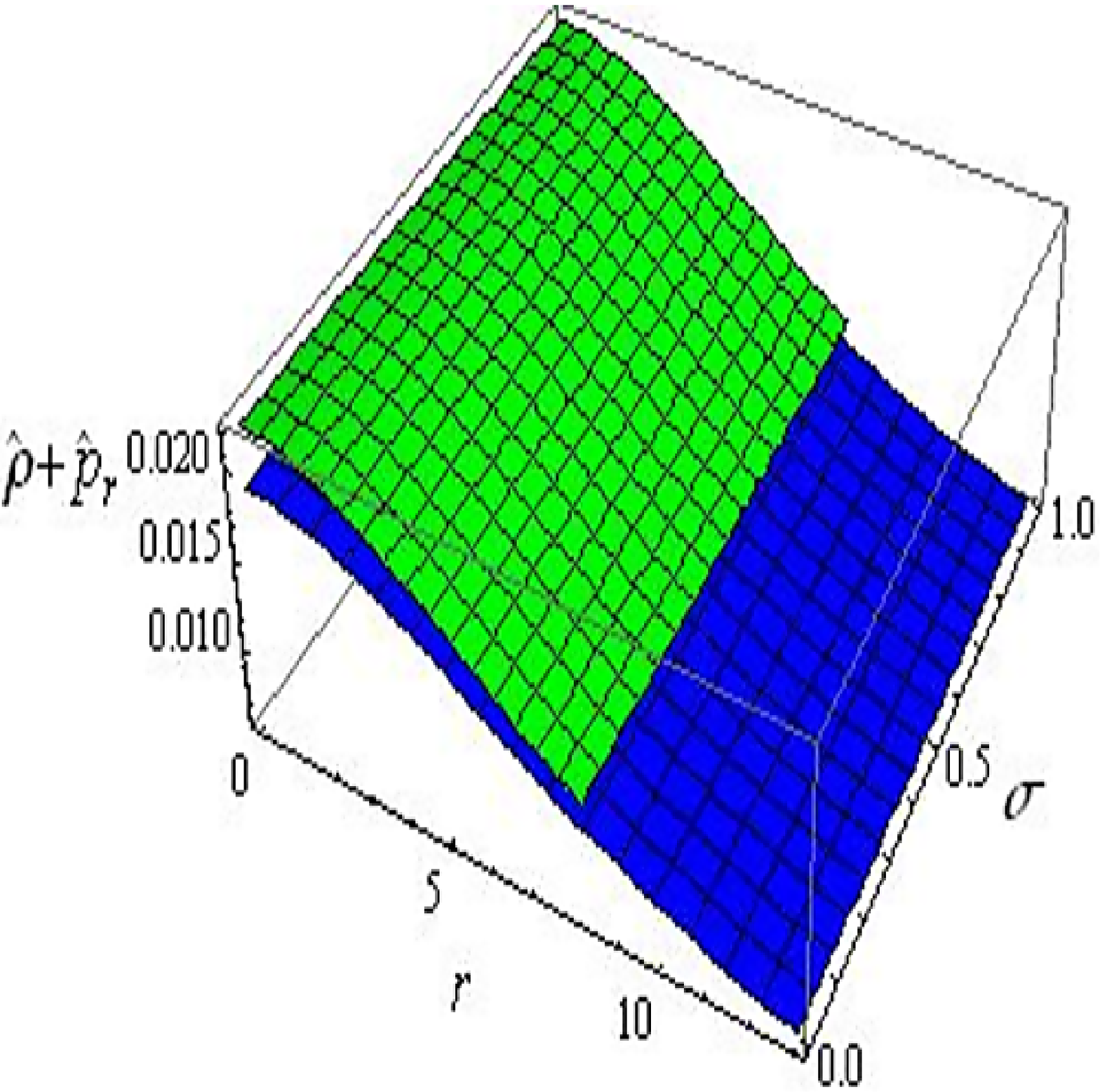,width=0.42\linewidth}
\epsfig{file=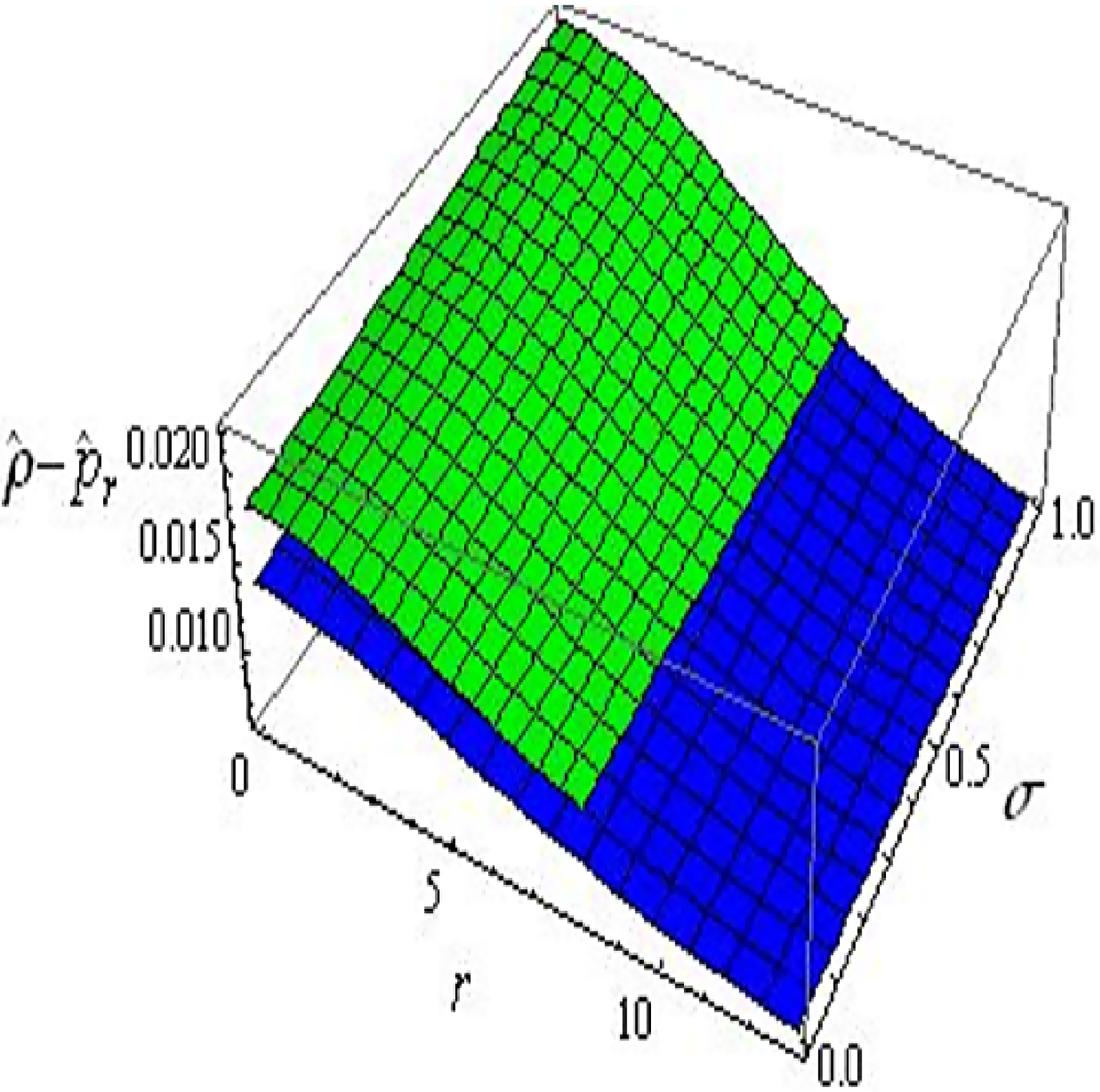,width=0.45\linewidth}
\epsfig{file=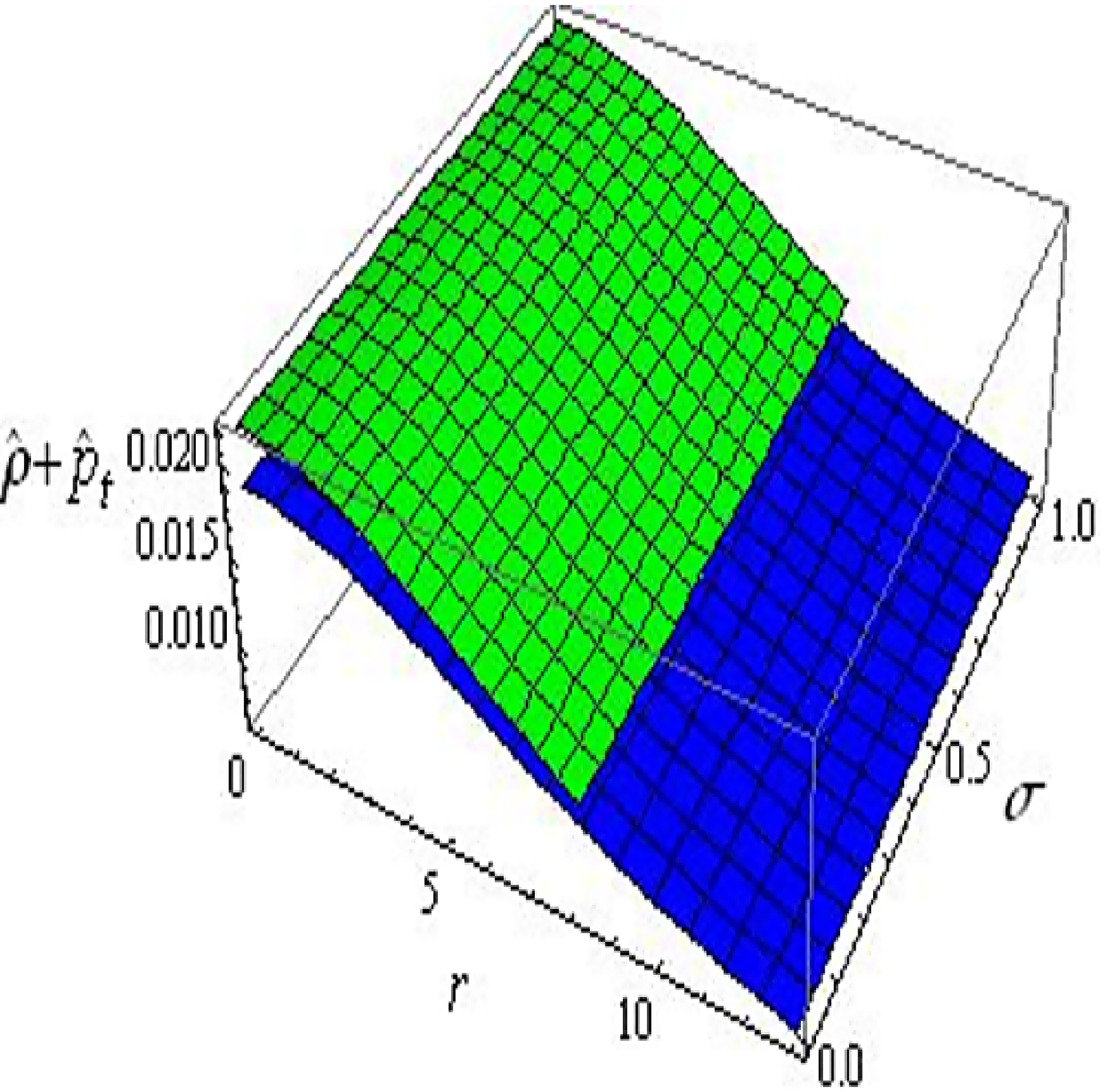,width=0.45\linewidth}
\epsfig{file=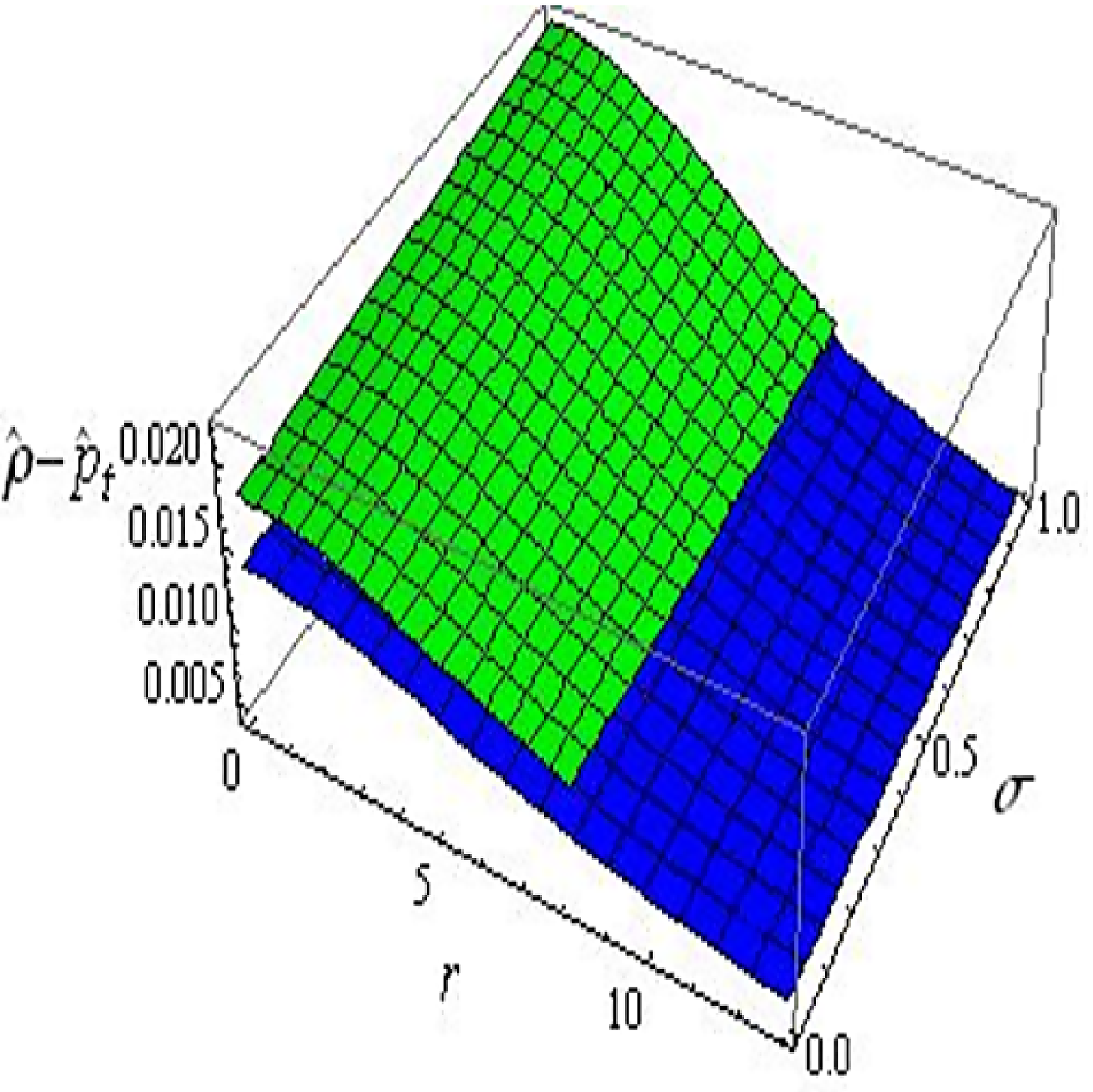,width=0.4\linewidth}
\epsfig{file=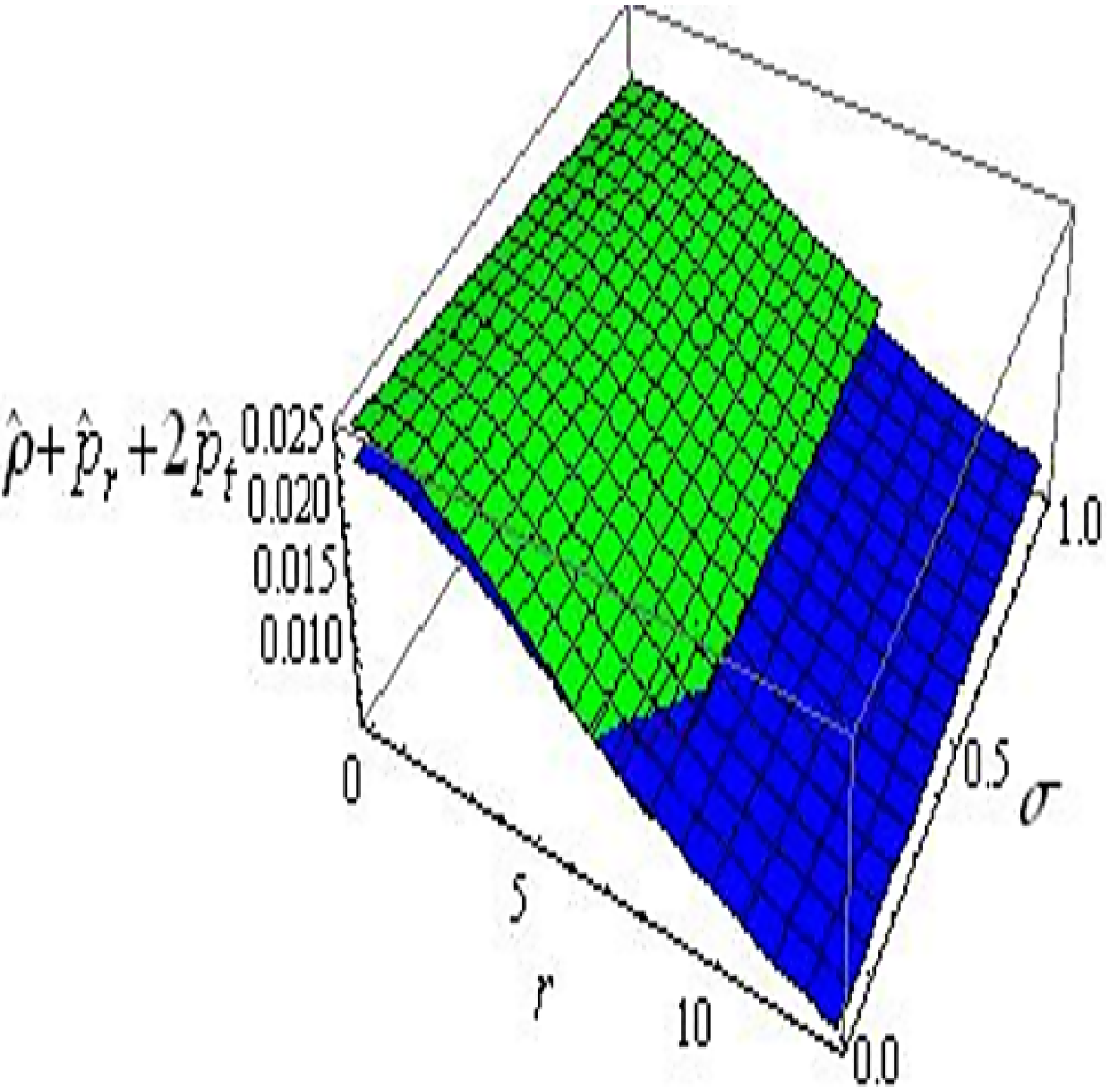,width=0.49\linewidth} \caption{Plots of energy
bounds for Her X-I (green) and PSR J 1416-2230 (blue) for solution
$\mathbf{I}$.}
\end{figure}

\subsection{Solution II}

In this section, we attain an alternative form of anisotropic
solution by employing a physically acceptable constraint on
$\vartheta_{0}^{0}$. The particular form of density-like constraint
is taken as
\begin{eqnarray}\label{61}
\vartheta_{0}^{0}&=&\rho.
\end{eqnarray}
With the help of  Eqs.(\ref{17}), (\ref{20}) and (\ref{21}), the
deformation functions take the form
\begin{eqnarray}\label{62}
g&=&\frac{r^2\left(\mathcal{A}^2+\mathcal{C}^2+r^2\right)}
{\mathcal{C}^2\left(\mathcal{A}^2+2 r^2\right)},\\
\nonumber h&=&\frac{1}{2}(-\frac{2
\ln\left(\mathcal{A}^2+r^2\right)}{\sigma }-\frac{2
\ln\left(\mathcal{A}^2+2 r^2\right)}{-1+\sigma
}+\left(\left(\mathcal{A}^2(-1+\sigma )+\mathcal{C}^2 (-1+3\sigma)
\right.\right.\\\nonumber&+&\left.\left.(-1+2 \sigma )\varrho\right)
\ln\left(\mathcal{C}^2 (1+\sigma )-\varrho+\left(\mathcal{A}^2+2
r^2\right)(-1+\sigma)\right)\right)\left(\varrho\sigma(-1+\sigma)
\right)^{-1}\\ \nonumber&+&\left(\left(\varrho(-1+2\sigma)
+\mathcal{C}^2 (1-3 \sigma )-\mathcal{A}^2 (-1+\sigma)\right)
\ln\left(\left(\mathcal{A}^2+2 r^2\right) (-1+\sigma )\right.\right.\\
\label{63}&+&\left.\left.\mathcal{C}^2
(1+\sigma)+\varrho\right)\right)\left((-1+\sigma)\sigma\varrho\right)^{-1}),
\end{eqnarray}
where
\begin{equation} \nonumber
\varrho=\sqrt{\mathcal{A}^4(-1+\sigma)^2+2\mathcal{A}^2
\mathcal{C}^2 (-1+\sigma)^2+\mathcal{C}^4 (1+\sigma)^2}.
\end{equation}
By using the same strategy as applied in solution \textbf{I}, the
matching conditions are computed as
\begin{eqnarray}\nonumber
\ln\left(1-\frac{2\mathcal{M}}{R}\right)&=&\ln\left(\mathcal{B}^2
(1+\frac{R^2}{\mathcal{A}^2})\right)+\frac{\sigma}{2}(-\frac{2
\ln\left(\mathcal{A}^2+r^2\right)}{\sigma }-\frac{2
\ln\left(\mathcal{A}^2+2 r^2\right)}{-1+\sigma
}\\\nonumber&+&\left(\left(\mathcal{A}^2(-1+\sigma )+\mathcal{C}^2
(-1+3\sigma)+(-1+2 \sigma )\varrho\right) \ln\left(\mathcal{C}^2
(1+\sigma )\right.\right.\\
\nonumber&-&\left.\left.\varrho+\left(\mathcal{A}^2+2
r^2\right)(-1+\sigma)\right)\right)\left(\varrho\sigma(-1+\sigma)
\right)^{-1}+\left(\left(\varrho(-1+2\sigma)\right.\right.\\
\nonumber&+&\left.\left.\mathcal{C}^2(1-3 \sigma)-\mathcal{A}^2
(-1+\sigma)\right)
\ln\left(\left(\mathcal{A}^2+2 r^2\right) (-1+\sigma )\right.\right.\\
\label{64}&+&\left.\left.\mathcal{C}^2
(1+\sigma)+\varrho\right)\right)\left((-1+\sigma)\sigma\varrho
\right)^{-1}),\\\label{65}
\mathcal{A}&=&\frac{\sqrt{-4\mathcal{C}^2\mathcal{M}
R^2+\mathcal{C}^2 R^3+R^5-\mathcal{C}^2R^3\sigma-R^5\sigma
}}{\sqrt{2\mathcal{C}^2\mathcal{M}-R^3+R^3\sigma}}.
\end{eqnarray}
The expressions of fluid parameters $(\hat{\rho}, \hat{p}_{r},
\hat{p}_{t})$ for anisotropic solution are evaluated as
\begin{eqnarray}\label{67}
\hat{\rho}&=&\frac{3\mathcal{A}^4+\mathcal{A}^2(3\mathcal{C}^2+7r^2)
+2r^2(\mathcal{C}^2+3r^2)}{\mathcal{C}^2(\mathcal{A}^2+2r^2)^2}(1-\sigma),
\\ \label{68} \hat{p}_{r}&=&\frac{\mathcal{C}^2-\mathcal{A}^2-3r^2}
{\mathcal{C}^2(\mathcal{A}^2+2r^2)}+\sigma(\frac{3\mathcal{A}^4
+\mathcal{A}^2(3\mathcal{C}^2+7r^2)
+2r^2(\mathcal{C}^2+3r^2)}{\mathcal{C}^2(\mathcal{A}^2+2r^2)^2}),\\
\nonumber
\hat{p}_{t}&=&\frac{\mathcal{C}^2-\mathcal{A}^2-3r^2}{\mathcal{C}^2
(\mathcal{A}^2+2r^2)}+\sigma(3\mathcal{A}^6+3\mathcal{A}^4\mathcal{C}^2
+8\mathcal{A}^4r^2-2\mathcal{A}^2\mathcal{C}^2r^2+18\mathcal{A}^2r^4+12
r^6\\\nonumber&+&\frac{\left(\mathcal{A}^2+2\mathcal{C}^2\right)r^2
\left(\mathcal{A}^6+\mathcal{A}^2
r^4-2\mathcal{C}^2r^4+\mathcal{A}^4
\left(\mathcal{C}^2+2r^2\right)\right)}{(\mathcal{C}-r)(\mathcal{C}+r)
\left(\mathcal{A}^2+r^2\right)+r^2
\left(\mathcal{A}^2+\mathcal{C}^2+r^2\right)\sigma})(\mathcal{C}^2
\left(\mathcal{A}^2+2 r^2\right)^3)^{-1}.\\ \label{69}
\end{eqnarray}
In this scenario, the anisotropic factor becomes
\begin{eqnarray}\nonumber
\hat{\Delta}&=&\left(\left(-\mathcal{A}^2+\mathcal{C}^2-3r^2\right)
\left(\mathcal{A}^2+2r^2\right)^2+\sigma\left(3\mathcal{A}^6
+3\mathcal{A}^4\mathcal{C}^2+8\mathcal{A}^4r^2-2
\mathcal{A}^2\mathcal{C}^2r^2\right.\right.\\
\nonumber&+&\left.\left.18\mathcal{A}^2
r^4+12r^6+\frac{\left(\mathcal{A}^2+2\mathcal{C}^2\right)r^2
\left(\mathcal{A}^6+\mathcal{A}^2r^4-2\mathcal{C}^2
r^4+\mathcal{A}^4\left(\mathcal{C}^2+2r^2\right)\right)}
{(\mathcal{C}-r)(\mathcal{C}+r)\left(\mathcal{A}^2+r^2\right)
+r^2\left(\mathcal{A}^2+\mathcal{C}^2
+r^2\right)\sigma}\right)\right)\\
\label{70}&\times&(\mathcal{C}^2
\left(\mathcal{A}^2+2r^2\right)^3)^{-1}.
\end{eqnarray}
We note that for $\sigma=0$, the anisotropic factor vanishes and our
model reduces to isotropic solution. The mass function, central
density and pressure take the form
\begin{eqnarray*}
m&=&-\frac{4 \pi  r^3 (\sigma -1)
\left(\mathcal{A}^2+\mathcal{C}^2+r^2\right)}{\mathcal{C}^2
\left(\mathcal{A}^2+2 r^2\right)},\\
\hat{\rho}_{c}&=&\frac{(1-\sigma ) \left(3 \mathcal{A}^4+3
\mathcal{A}^2 \mathcal{C}^2\right)}{\mathcal{A}^4 \mathcal{C}^2},\\
\hat{p}_{rc}&=&\frac{3 \sigma +1}{\mathcal{A}^2}+\frac{3 \sigma
-1}{\mathcal{C}^2}.
\end{eqnarray*}
\begin{figure}\center
\epsfig{file=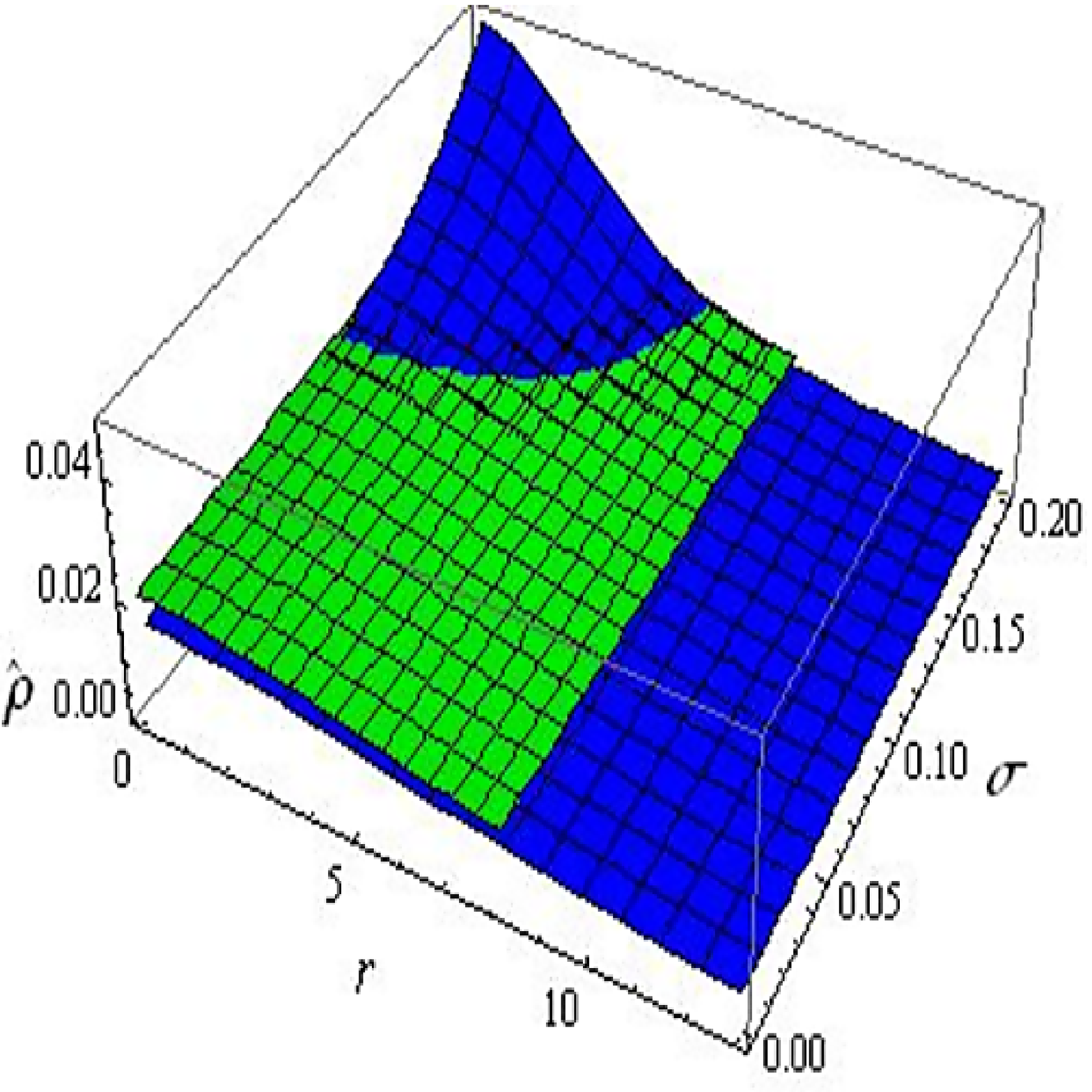,width=0.45\linewidth}\epsfig{file=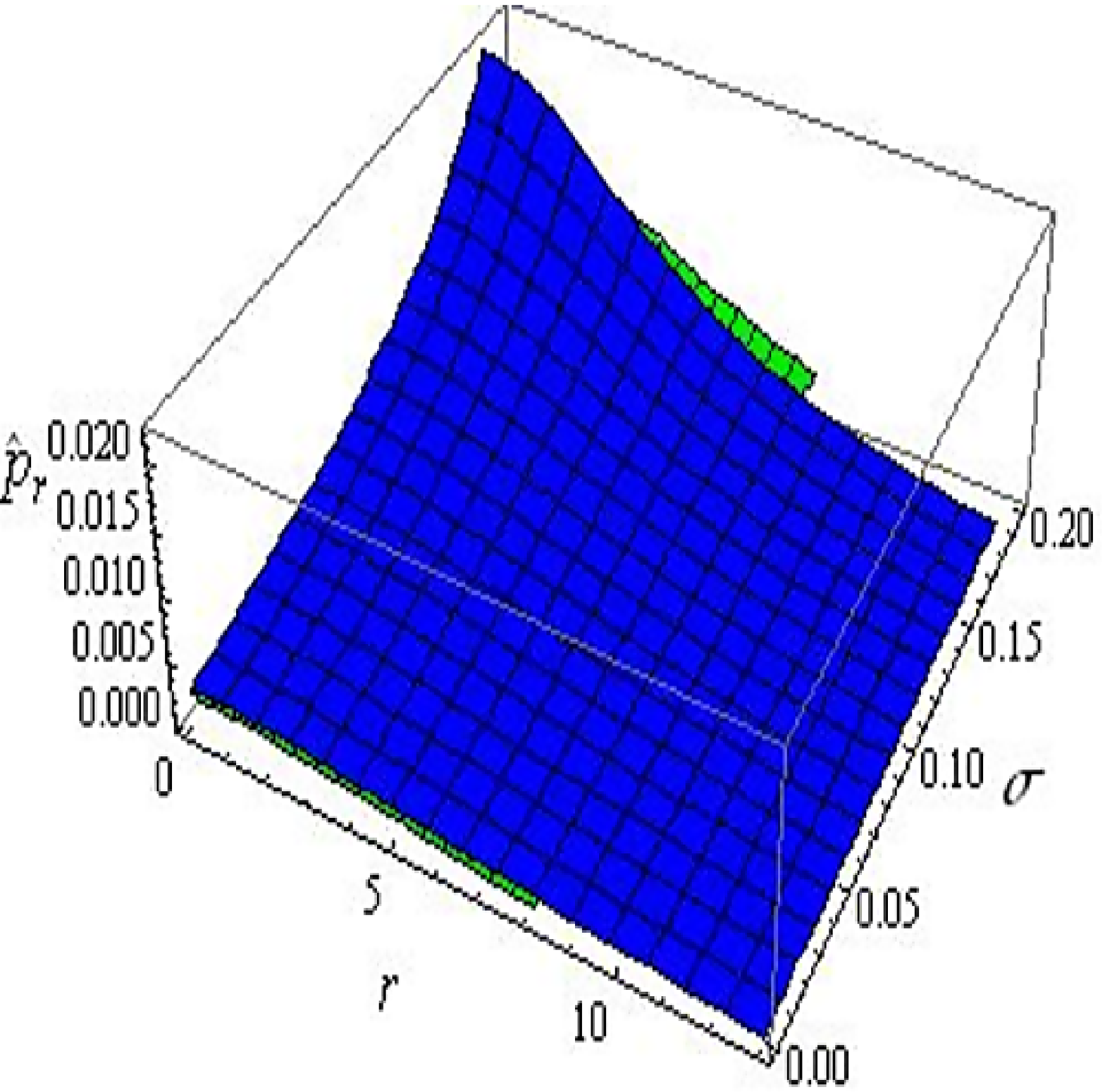,
width=0.45\linewidth} \epsfig{file=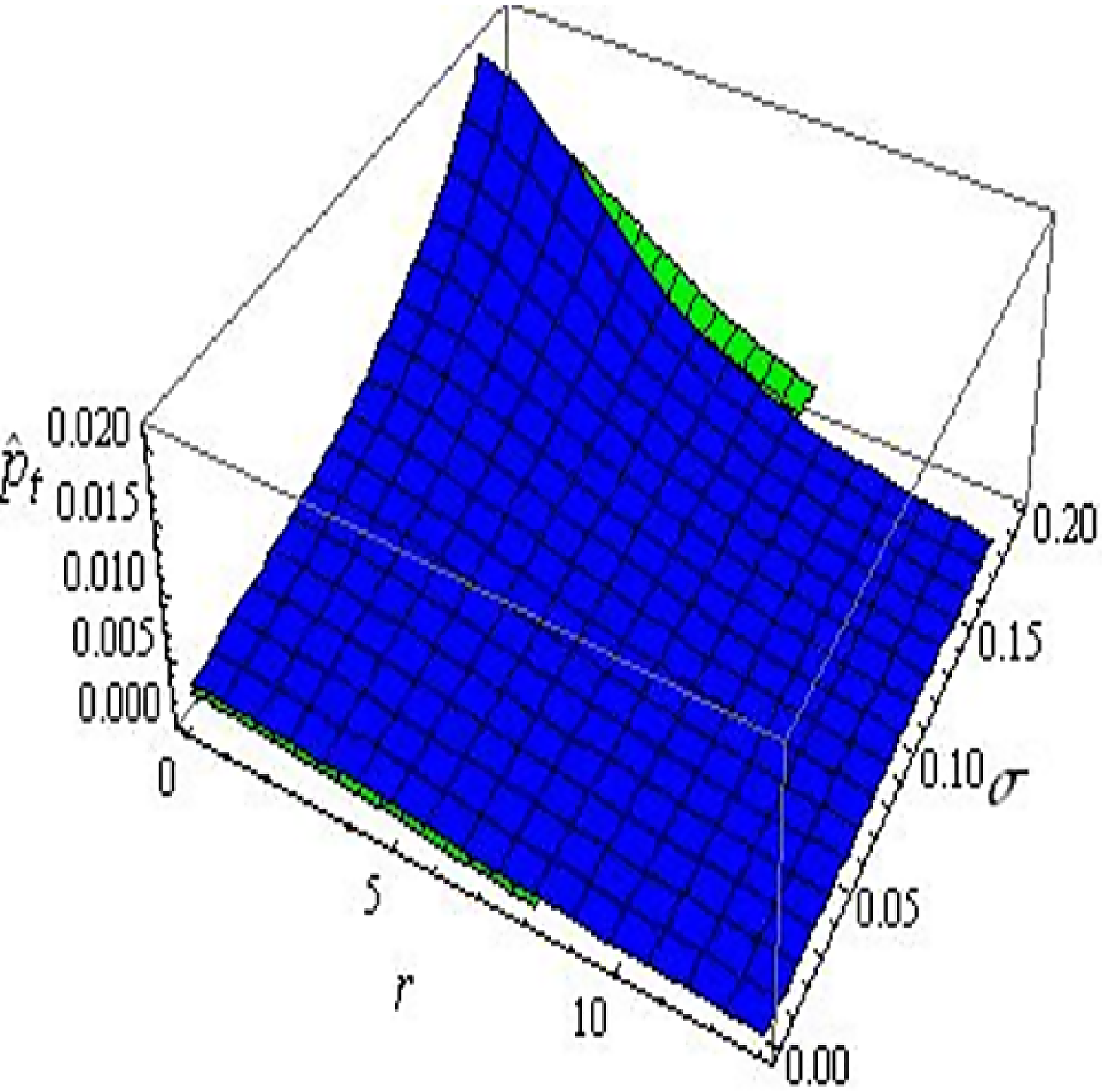,width=0.45\linewidth}
\epsfig{file=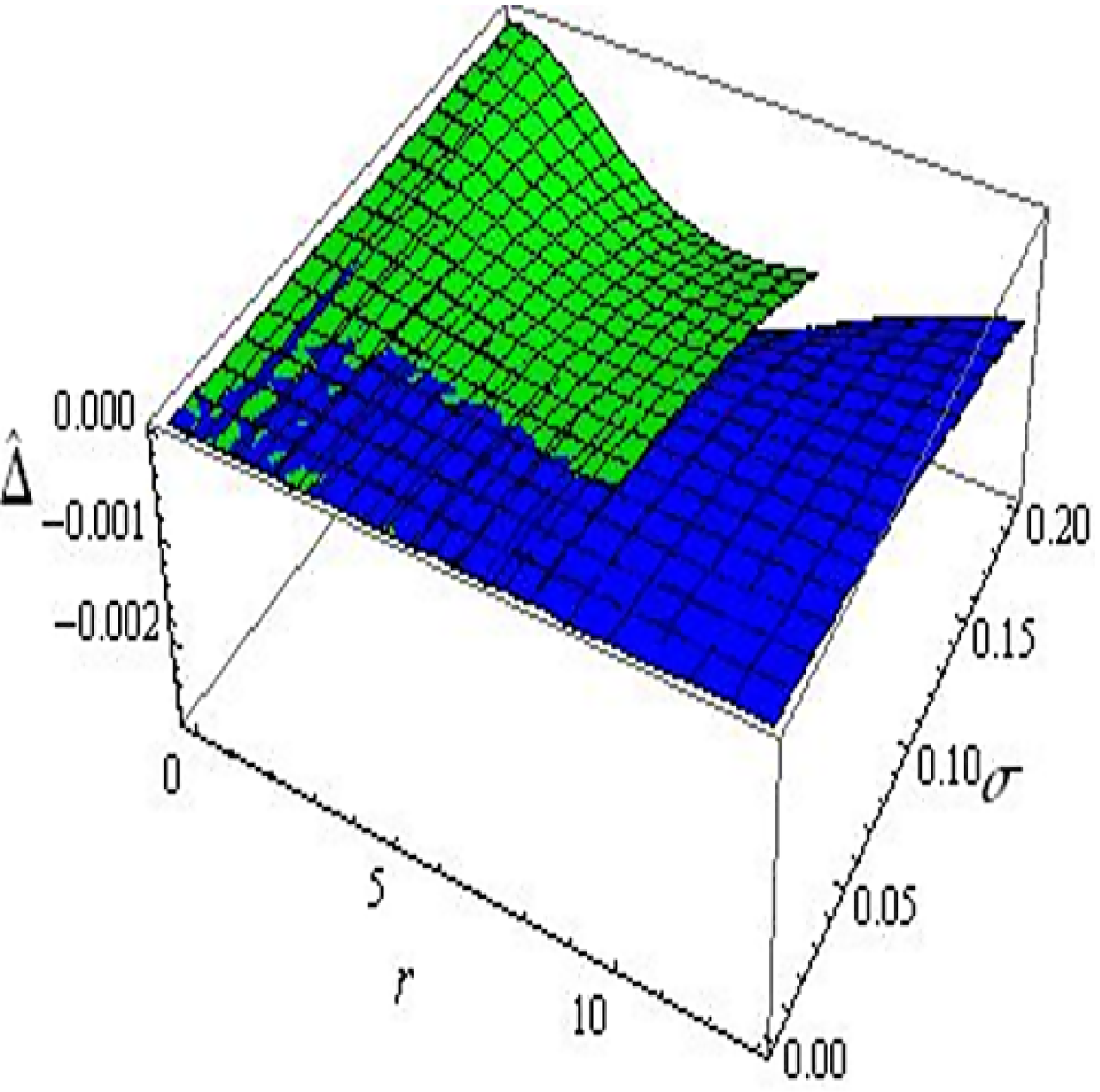,width=0.45\linewidth} \caption{Plots of
$\hat{\rho}$, $\hat{p}_{r}$, $\hat{p}_{t}$ and $\hat{\Delta}$ for
Her X-I (green) and PSR J 1416-2230 (blue) for solution
$\mathbf{II}$.}
\end{figure}
\begin{figure}\center
\epsfig{file=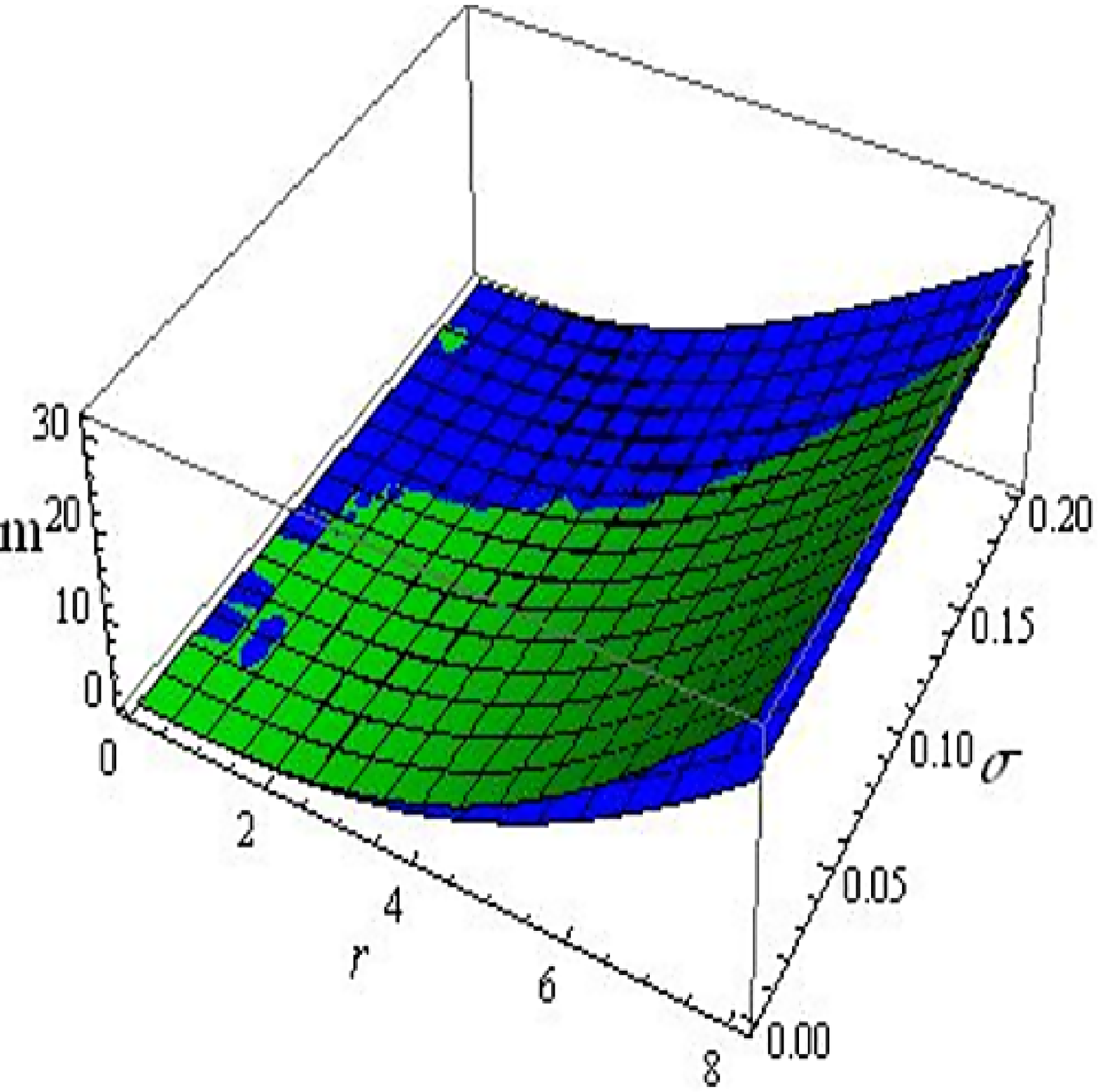,width=0.45\linewidth} \caption{Plot of mass for
Her X-I (green) and PSR J 1416-2230 (blue) for solution
$\mathbf{II}$.}
\end{figure}
\begin{figure}\center
\epsfig{file=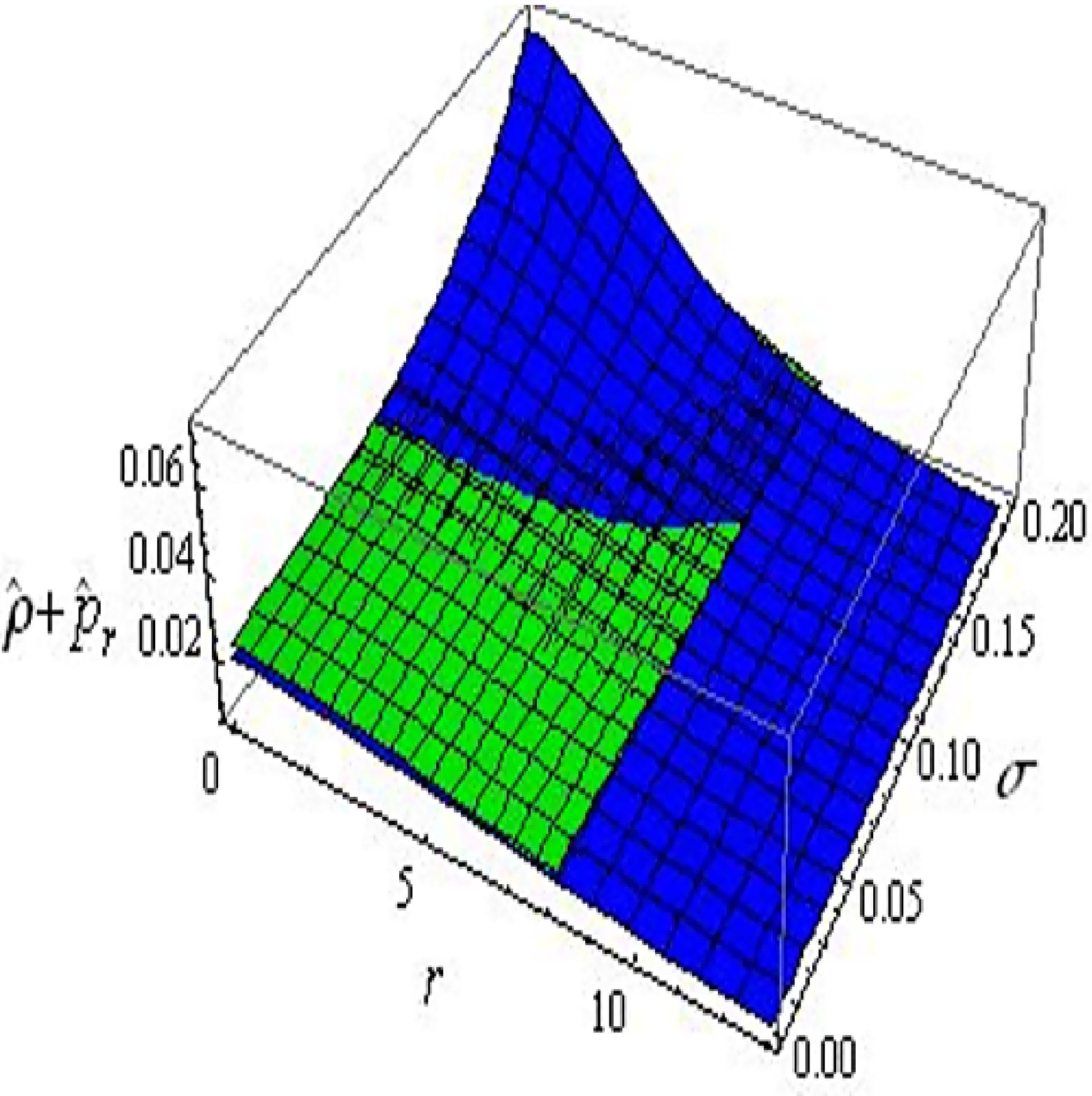,width=0.45\linewidth}\epsfig{file=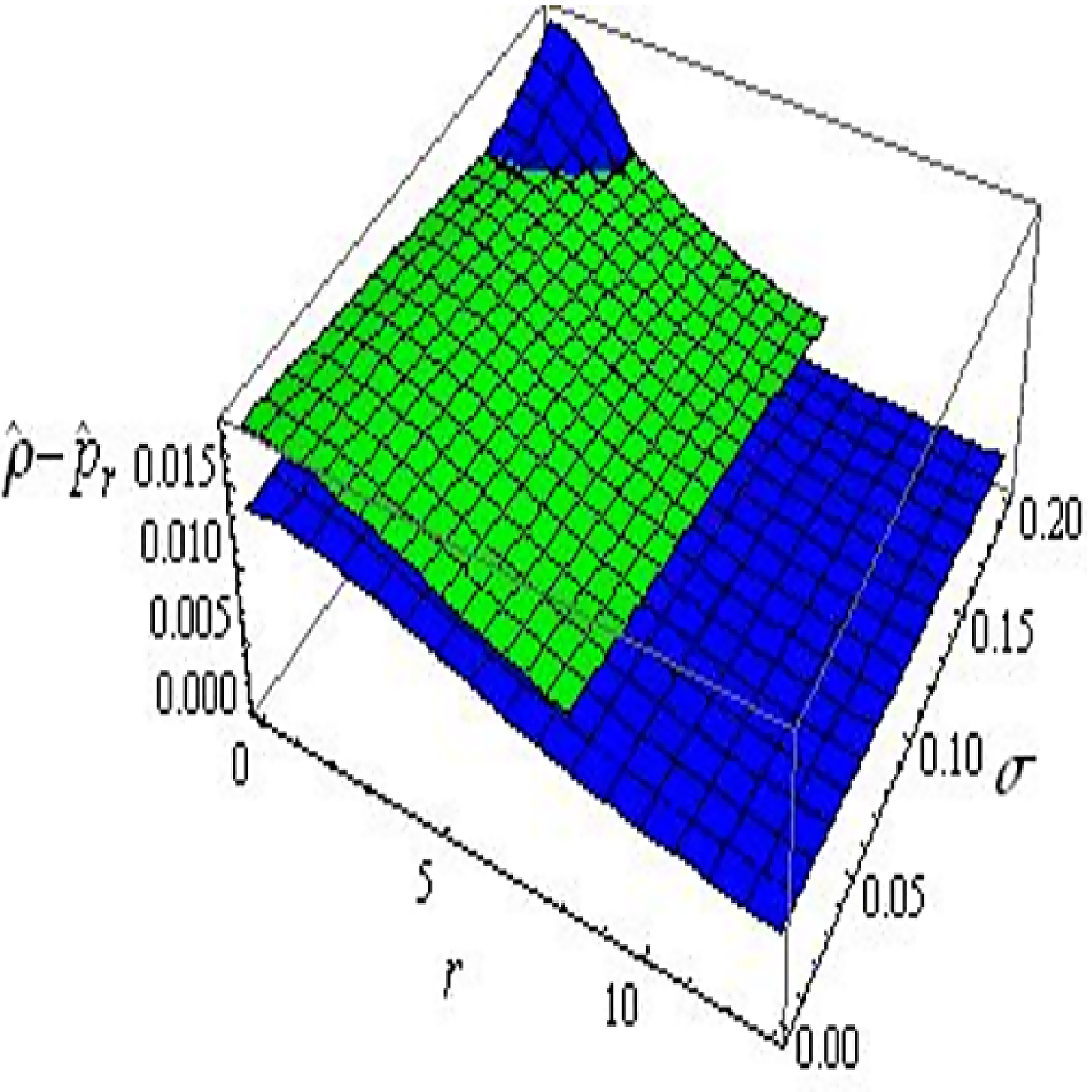,
width=0.45\linewidth} \epsfig{file=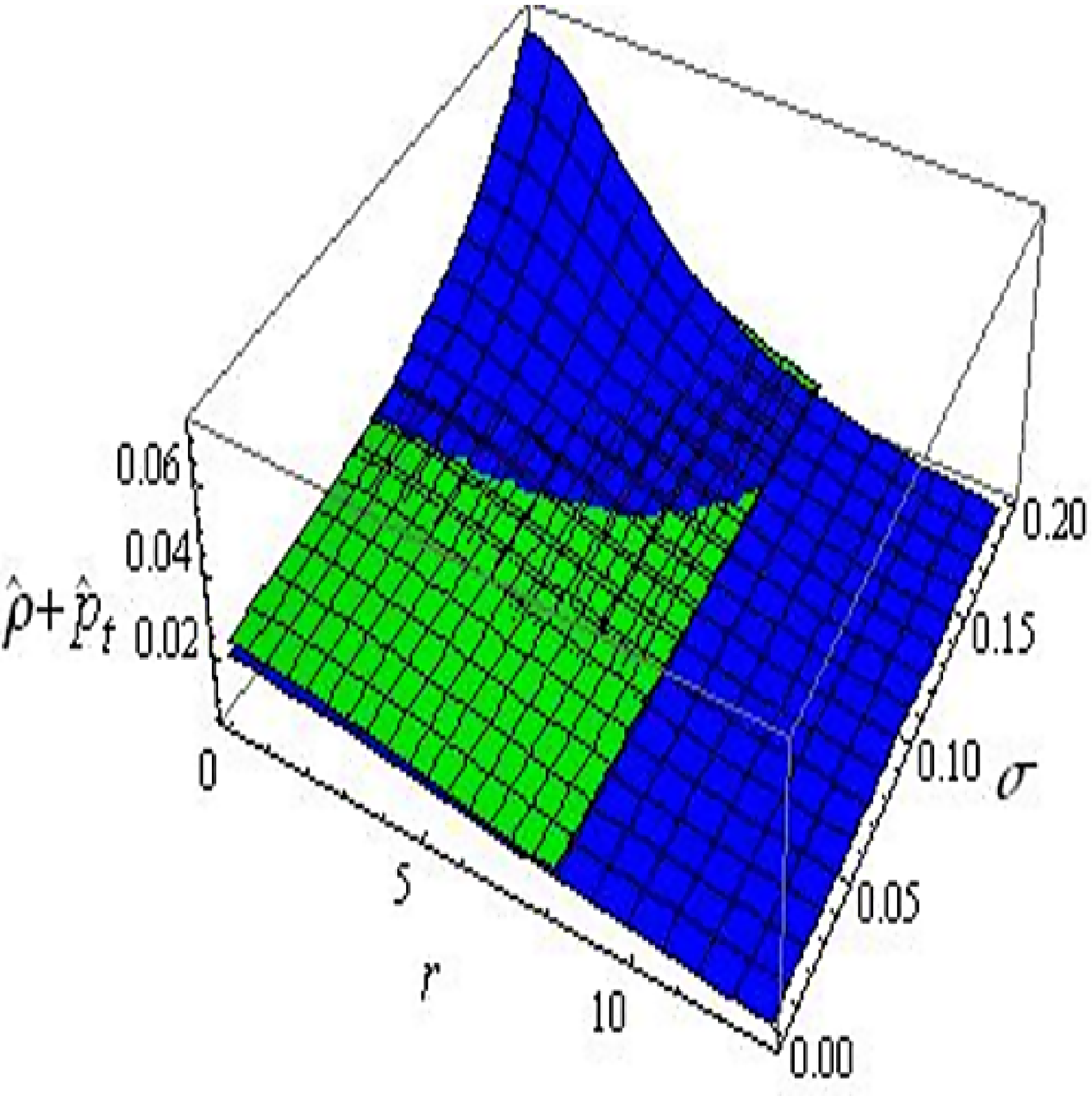,width=0.45\linewidth}
\epsfig{file=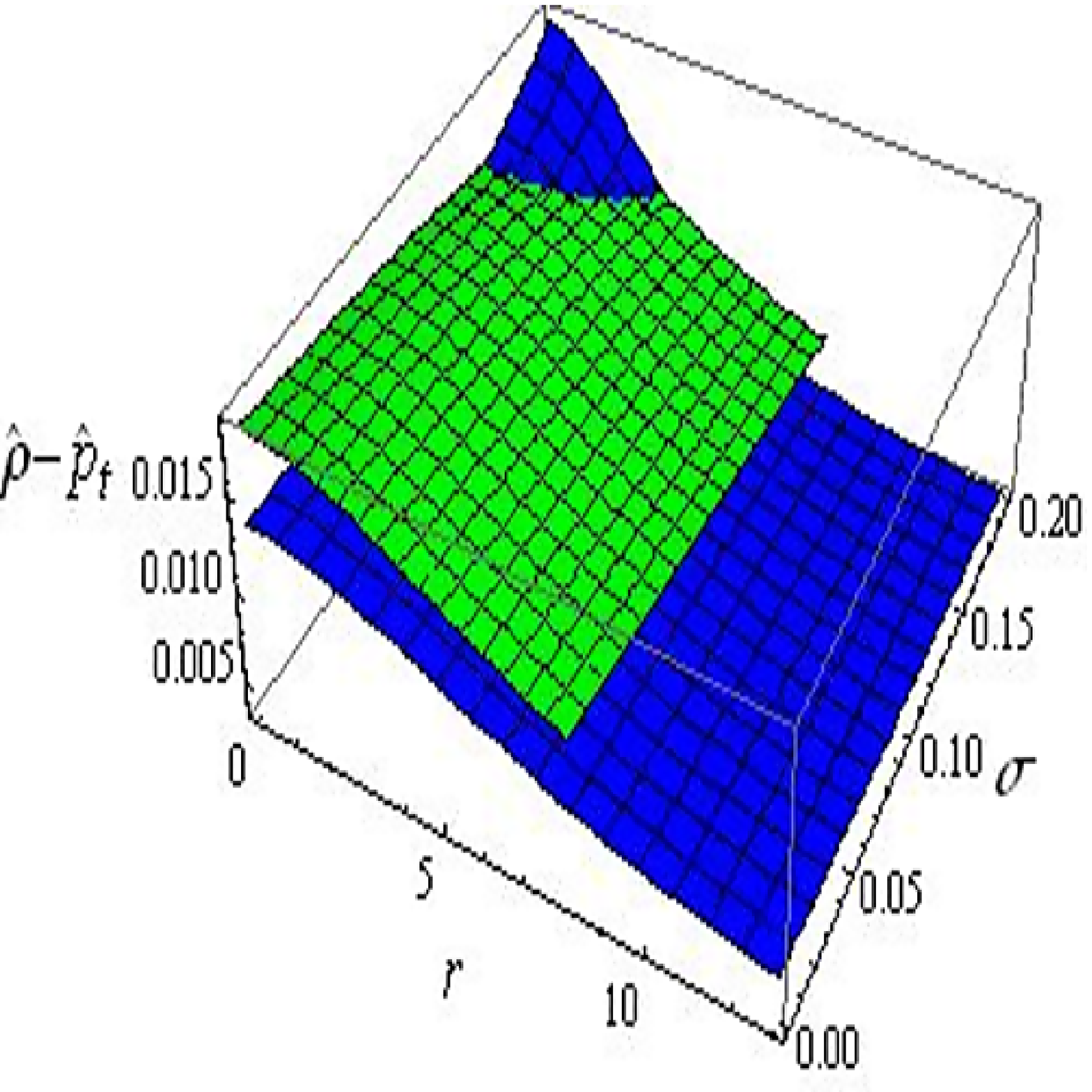,width=0.45\linewidth}
\epsfig{file=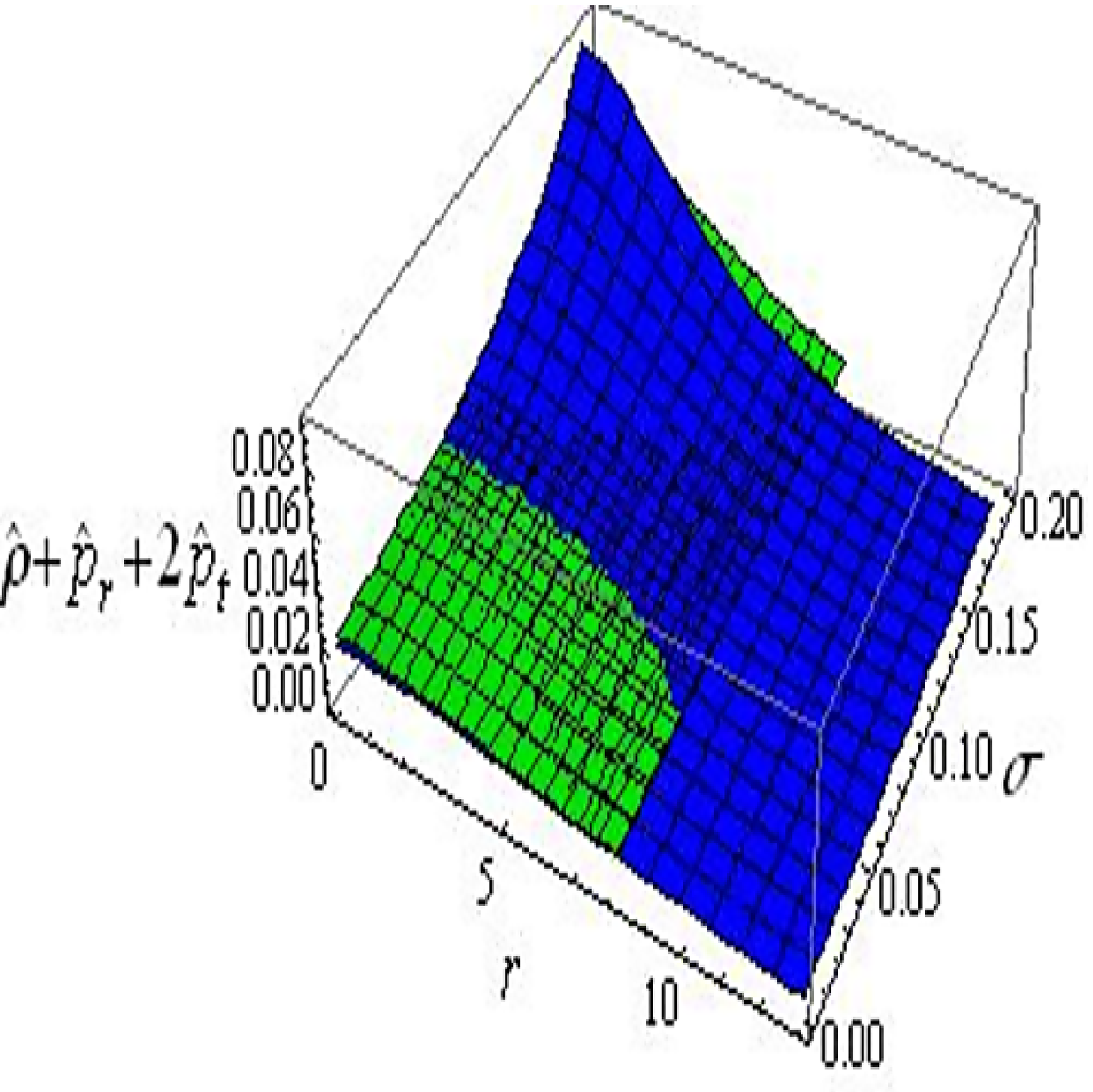,width=0.45\linewidth} \caption{Plots of energy
bounds for Her X-I (green) and PSR J 1416-2230 (blue) for solution
$\mathbf{II}$.}
\end{figure}

To determine the physical characteristics of solution $\textbf{II}$,
the expressions of $\mathcal{C}$ and $\mathcal{A}$ are taken from
Eqs.(\ref{44}) and (\ref{65}), respectively. The graphical behavior
of matter variables ($\hat{\rho}, \hat{p}_{r}, \hat{p}_{t}$ and
$\hat{\Delta}$) for the stars Her X-I ($\mathcal{M}=1.25375
\text{km}$, $R=8.10 \text{km}$) and PSR J 1416-2230
($\mathcal{M}=2.90575 \text{km}$, $R=13 \text{km}$) are shown in
Figure \textbf{4}. It is found that the density as well as
anisotropic pressures follow the pattern of solution \textbf{I},
i.e., they are finite, monotonically decreasing as well as regular.
However, the impact of anisotropic factor is found to be negative
which indicates that the anisotropic force leads the system towards
more compact matter distribution. It is noted that this solution
satisfies the viability criterion for smaller choices of $\sigma$ as
compared to solution \textbf{I}. For larger values of decoupling
parameter, the plot of tangential pressure becomes negative which
exhibits the non-realistic behavior of the stellar structure. The
consistency of mass function and energy bounds are displayed in
Figures \textbf{5} and \textbf{6}, respectively, which represent
that the model fulfills all criteria of being physically acceptable
solution.

\section{Dynamical Equilibrium and Stability Constraints}

This section investigates to determine the equilibrium state through
Tolman-Oppenheimer-Volkoff (TOV) equation as well as to perform the
stability criteria via speed of sound constraint and adiabatic index
for the developed anisotropic models.

\subsection{TOV Equation for Gravitationally Decoupled Models}

Tolman \cite{23a}, Oppenheimer and Volkoff \cite{23b} proposed that
the sum of all physical forces, viz. hydrostatic force $(f_{h})$,
gravitational force $(f_{g})$ and anisotropic force $(f_{a})$ must
be zero to maintain the system into an equilibrium state, i.e.,
\begin{equation}
f_{h}+f_{g}+f_{a}=0.
\end{equation}
For our spacetime, the TOV equation takes the from
\begin{equation}
-[p^{'}+\sigma\vartheta_{1}^{1'}]-[\frac{\xi^{'}}{2}(\rho+p)+\frac{\sigma
h^{'}}{2}(\rho+p)+\sigma
\frac{\eta^{'}}{2}(\vartheta_{1}^{1}-\vartheta_{0}^{0})]
+[\frac{2\sigma}{r}(\vartheta_{2}^{2}-\vartheta_{1}^{1})]=0.
\end{equation}
The factors in square brackets denote the forces $f_{h}$, $f_{g}$
and $f_{a}$, respectively which explicitly can be written as
\begin{eqnarray}
f_{h}&=&-(p^{'}+\sigma\vartheta_{1}^{1'}),\\
f_{g}&=&-(\frac{\xi^{'}}{2}(\rho+p)+\frac{\sigma
h^{'}}{2}(\rho+p)+\sigma\frac{\eta^{'}}{2}(\vartheta_{1}^{1}-\vartheta_{0}^{0})),\\
f_{a}&=&-\frac{2\sigma}{r}(\vartheta_{1}^{1}-\vartheta_{2}^{2}).
\end{eqnarray}
The profile of these forces corresponding to the solutions
$\textbf{I}$ and $\textbf{II}$ are plotted in Figures $\textbf{7}$
and $\textbf{8}$, respectively. The figures show the dominating
behavior of the gravitational force which is counter-balanced by the
effects of anisotropic as well as hydrostatic forces. Hence, our
proposed models are in equilibrium state under the combine influence
of these fundamental forces.
\begin{figure}\center
\epsfig{file=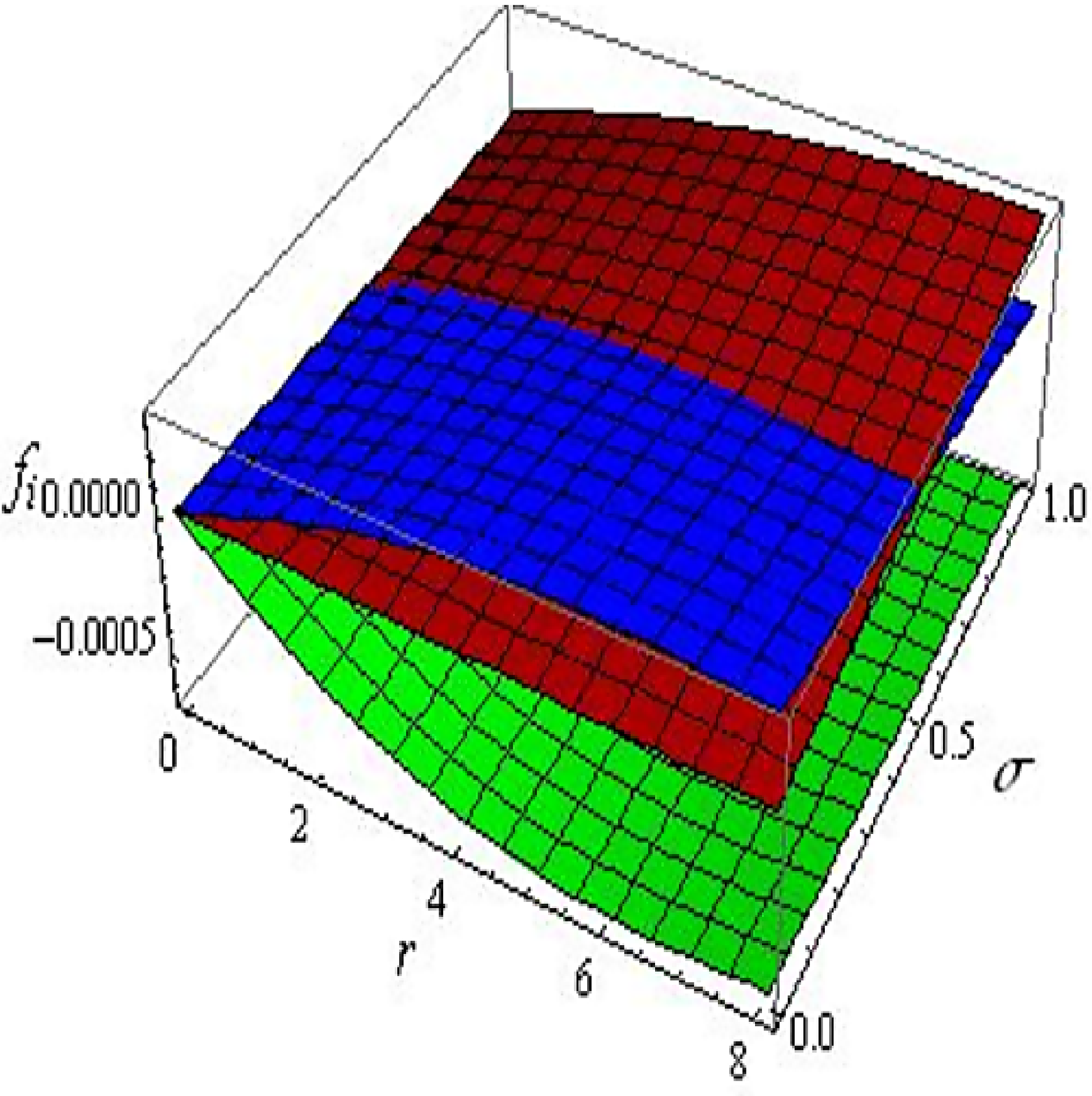,width=0.45\linewidth}
\epsfig{file=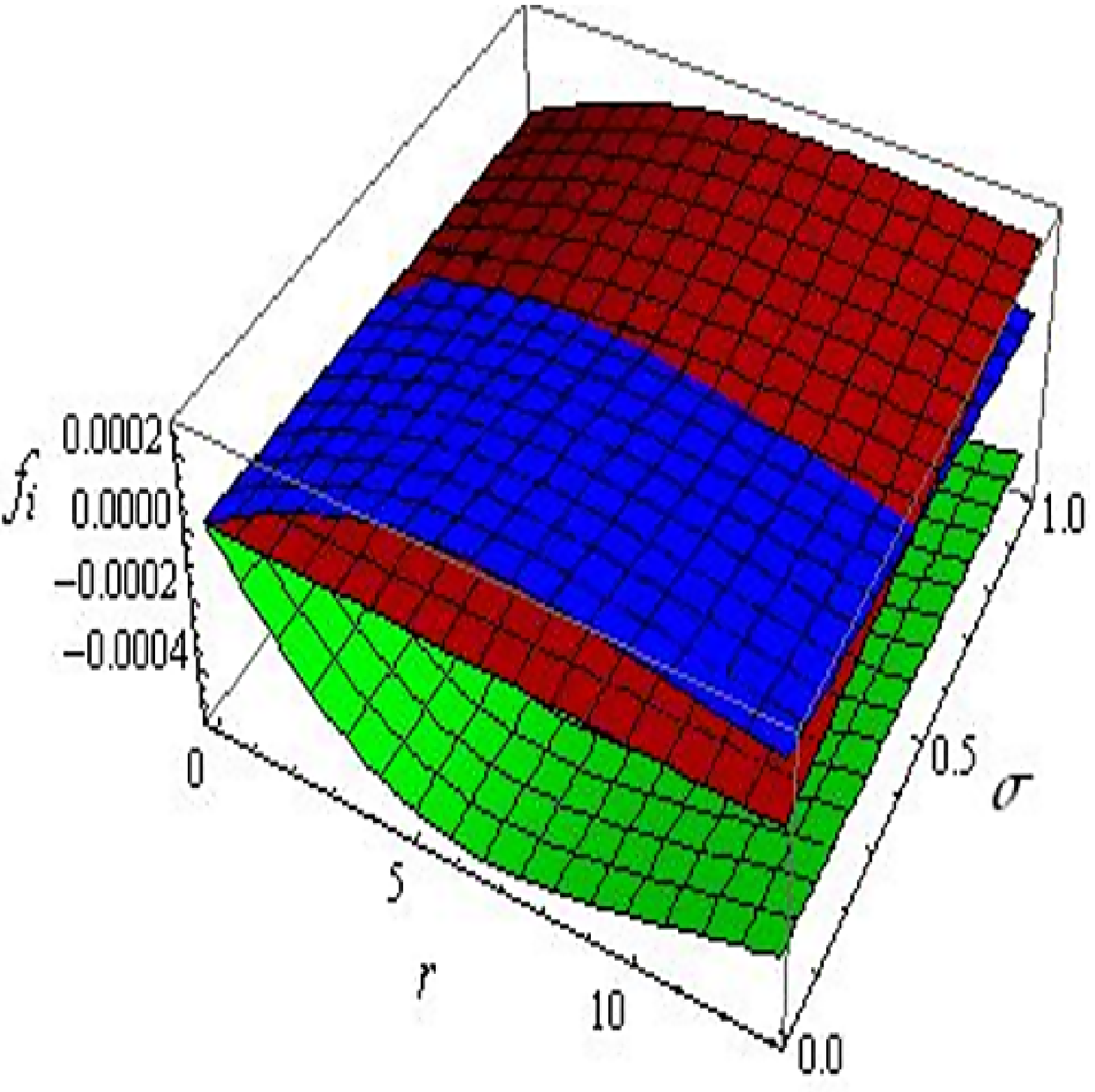,width=0.49\linewidth} \caption{Plots of
fundamental forces ($f_{g}$ (green), $f_{a}$ (red) and $f_{h}$
(blue)) for Her X-I (left) and PSR J 1416-2230 (right) corresponding
to solution $\mathbf{I}$.}
\end{figure}
\begin{figure}\center
\epsfig{file=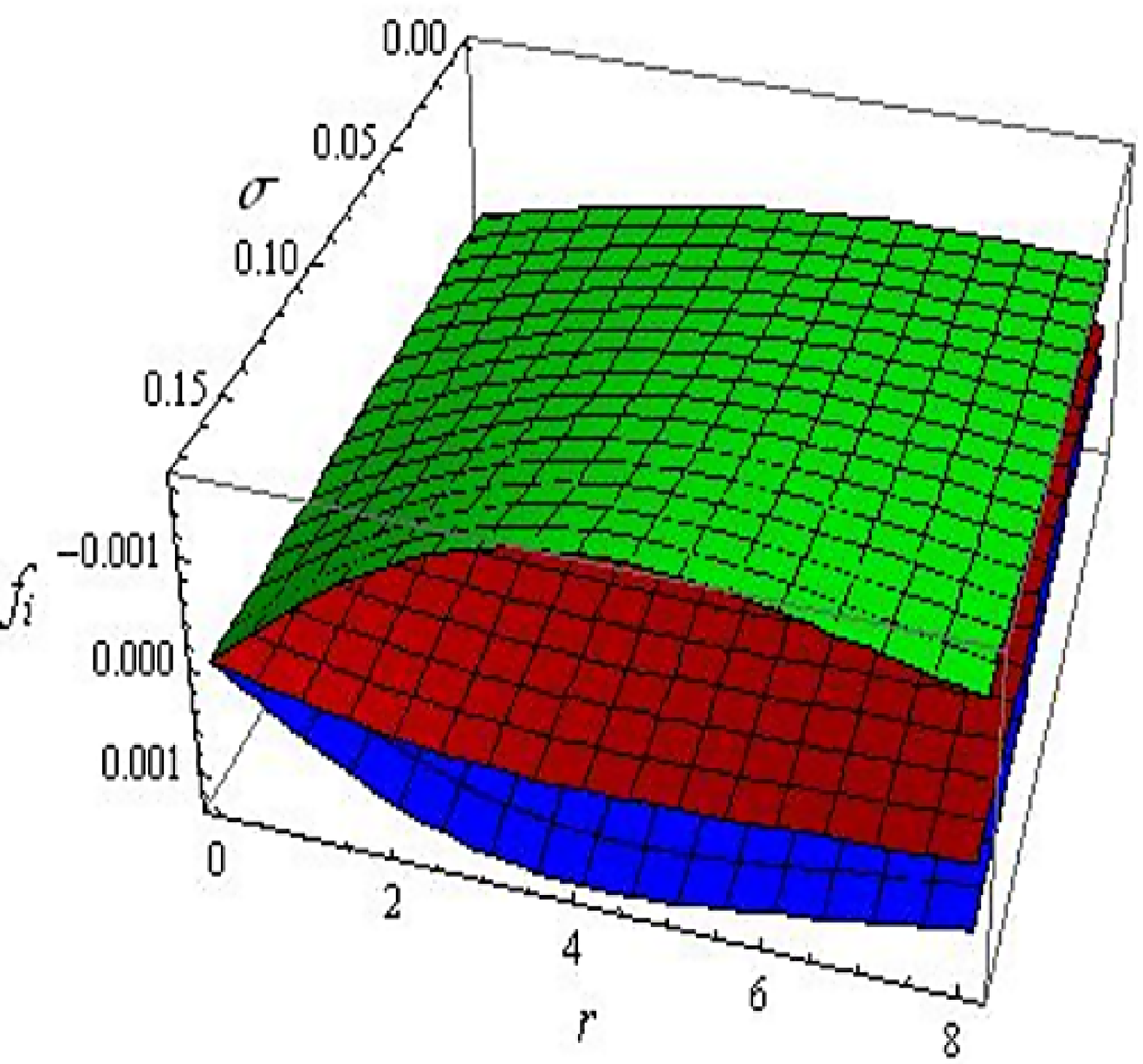,width=0.45\linewidth}
\epsfig{file=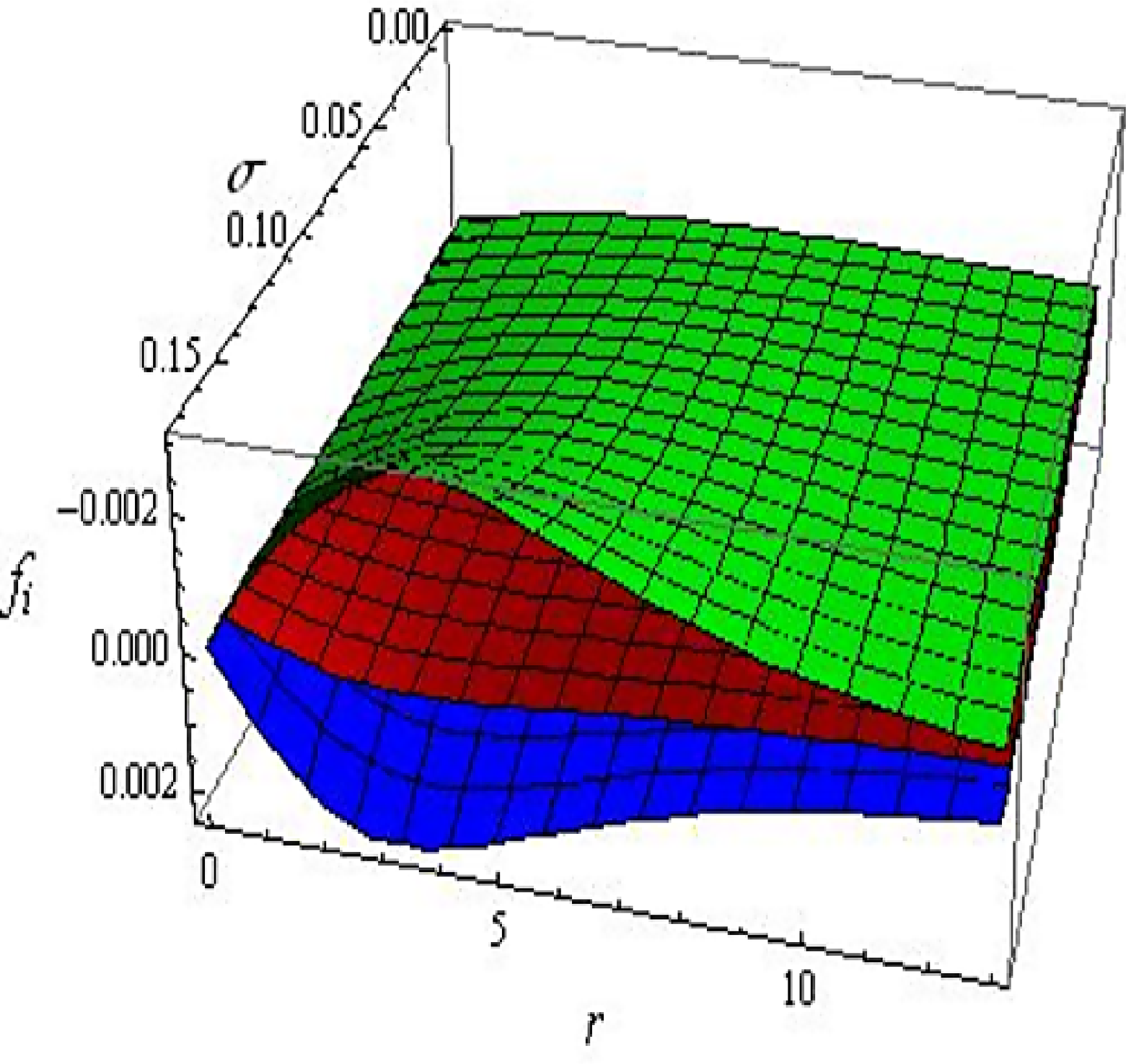,width=0.49\linewidth} \caption{Plots of
fundamental forces ($f_{g}$ (green), $f_{a}$ (red) and $f_{h}$
(blue)) for Her X-I (left) and PSR J 1416-2230 (right) corresponding
to solution $\mathbf{II}$.}
\end{figure}

\subsection{Causality Condition}

Once the system has achieved its equilibrium stage then the next
question arises whether it is stable or not. To do so, we will study
the stability of the anisotropic models through Abreu et al.
\cite{24a} technique based on Herrera's cracking concept \cite{24b}
which demands
\begin{eqnarray}\nonumber
\left\{\begin{array}{lll} -1\leq v_{st}^{2}-v_{sr}^{2}\leq
0\Rightarrow \text{Potentially stable},\\ 0<
v_{st}^{2}-v_{sr}^{2}\leq 1\Rightarrow \text{Potentially unstable},
\end{array}\right.
\end{eqnarray}
where
\begin{eqnarray}
v_{st}^{2}=\frac{d\hat{p}_{t}}{d\hat{\rho}}, \quad
v_{sr}^{2}&=&\frac{d\hat{p}_{r}}{d\hat{\rho}}.
\end{eqnarray}
The above conditions can be unified as $|v_{st}^{2}-v_{sr}^{2}|\leq
1$ \cite{24c} which proposed that ``no cracking" concept is
necessary for the potentially stable regions. This procedure
requires that for an physically acceptable model the transverse as
well as radial sound velocities should be less than 1, i.e.,
$v_{st}^{2},~ v_{sr}^{2}<1$ which are also known as causality
conditions. It is clear from Figures \textbf{9} and \textbf{10} that
both anisotropic models meet the causality constraints along with
the no cracking condition. Thus, the stability criteria are
satisfied for both constructed solutions.
\begin{figure}\center
\epsfig{file=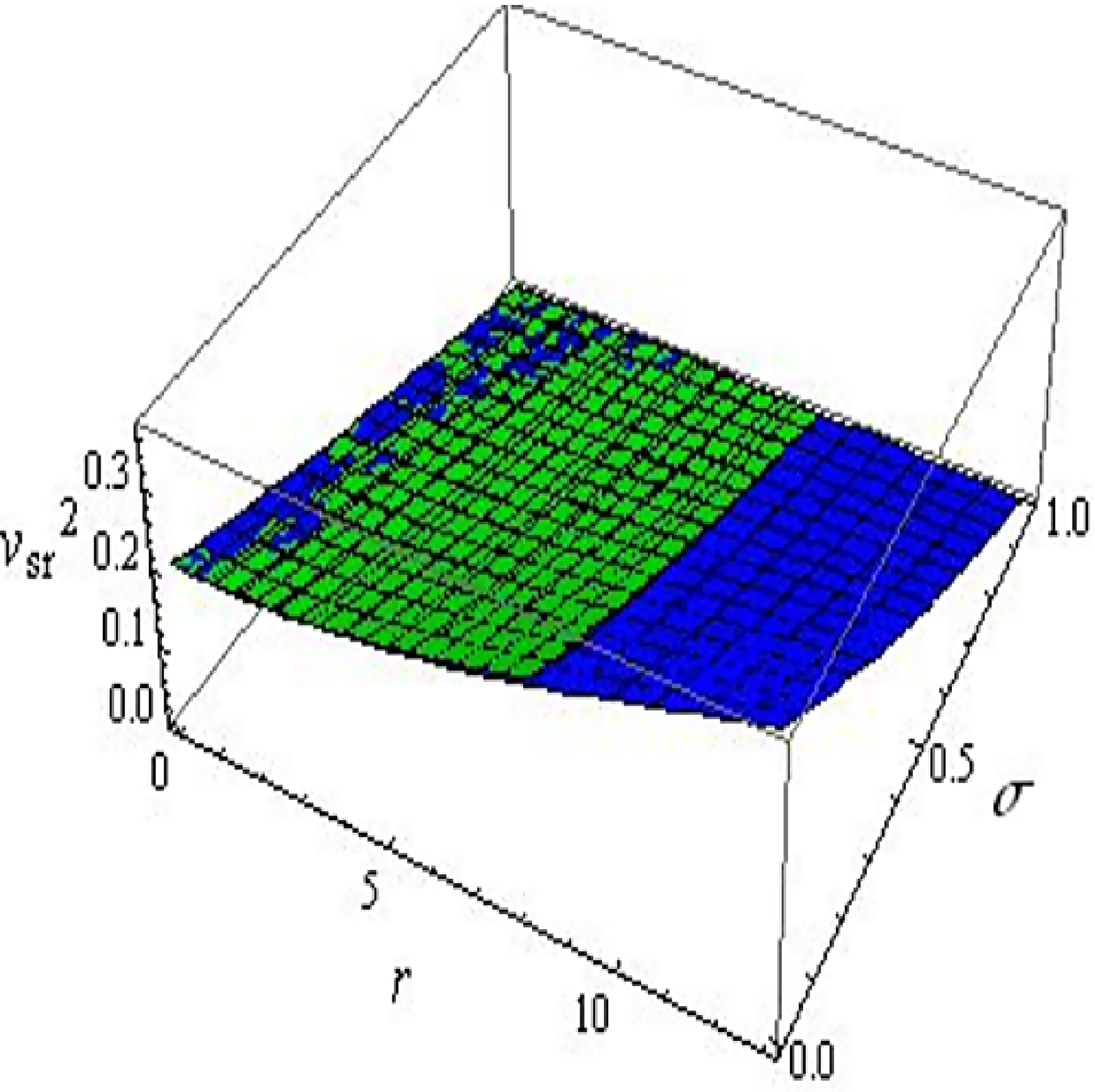, width=0.45\linewidth}\epsfig{file=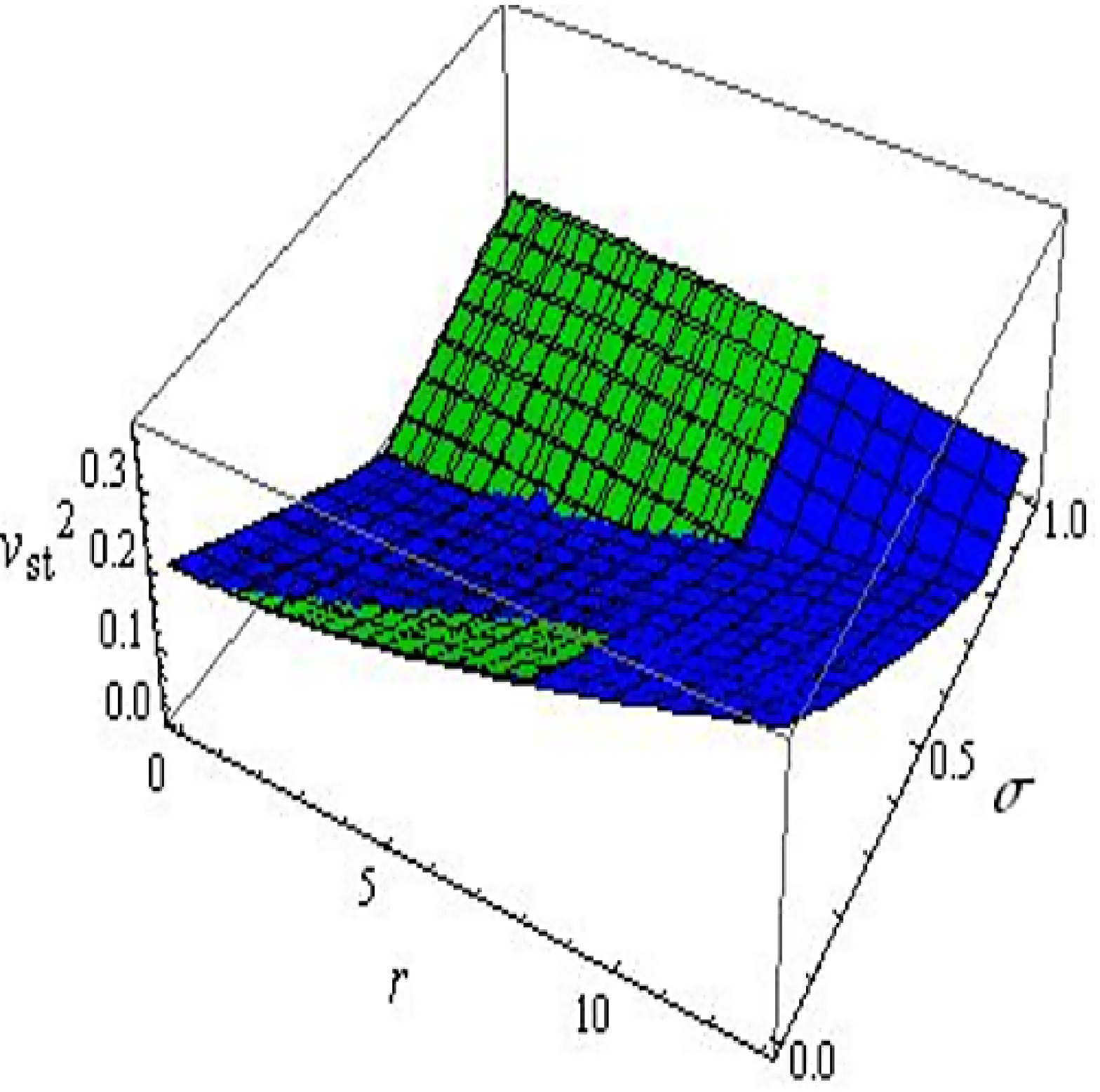,
width=0.45\linewidth}\\ \epsfig{file=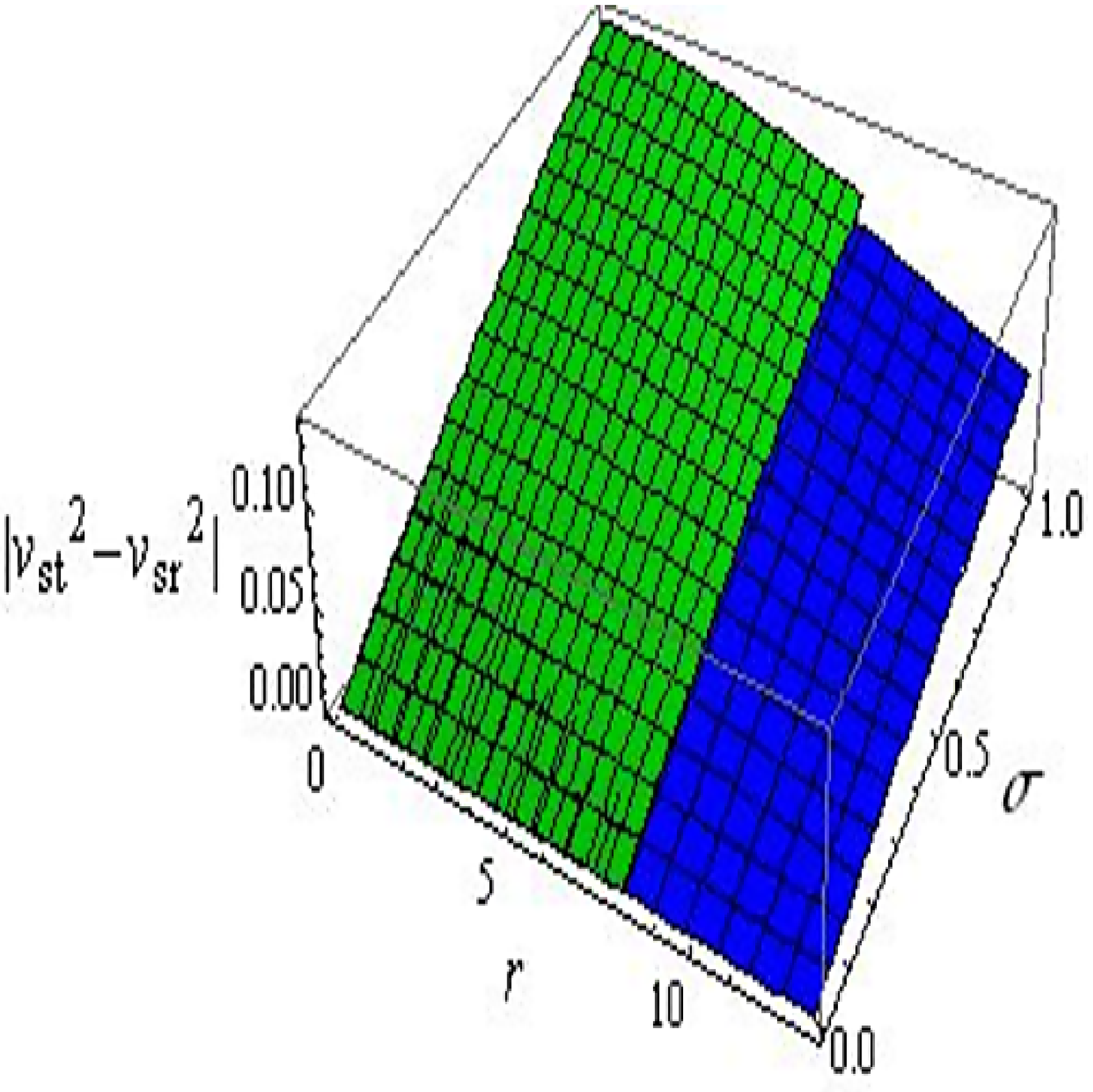, width=0.45\linewidth}
\caption{Plots of $v_{sr}^{2}$, $v_{st}^{2}$ and
$|v_{st}^{2}-v_{sr}^{2}|$ for Her X-I (green) and PSR J 1416-2230
(blue) corresponding to solution $\mathbf{I}$.}
\end{figure}
\begin{figure}\center
\epsfig{file=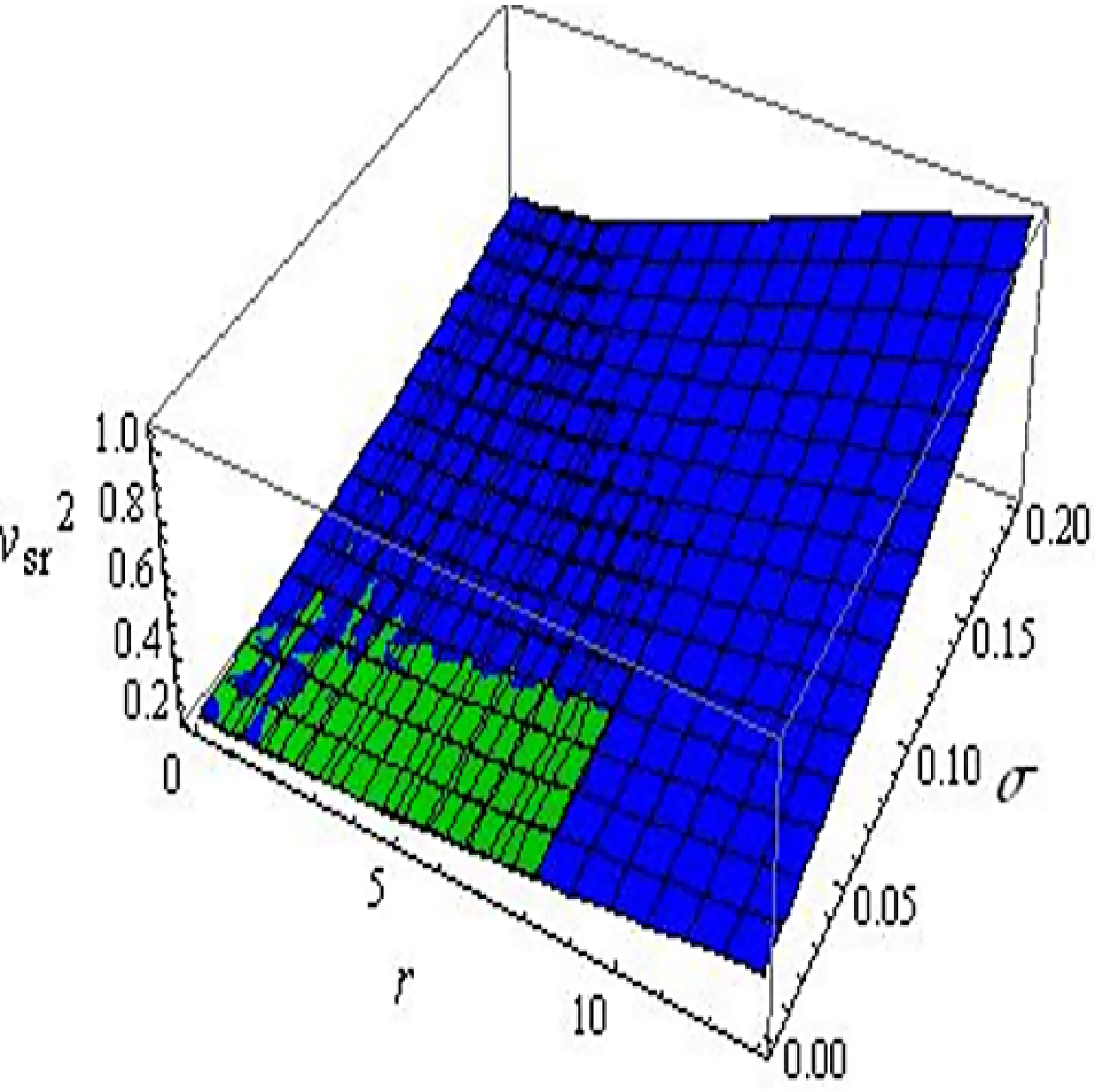, width=0.45\linewidth}\epsfig{file=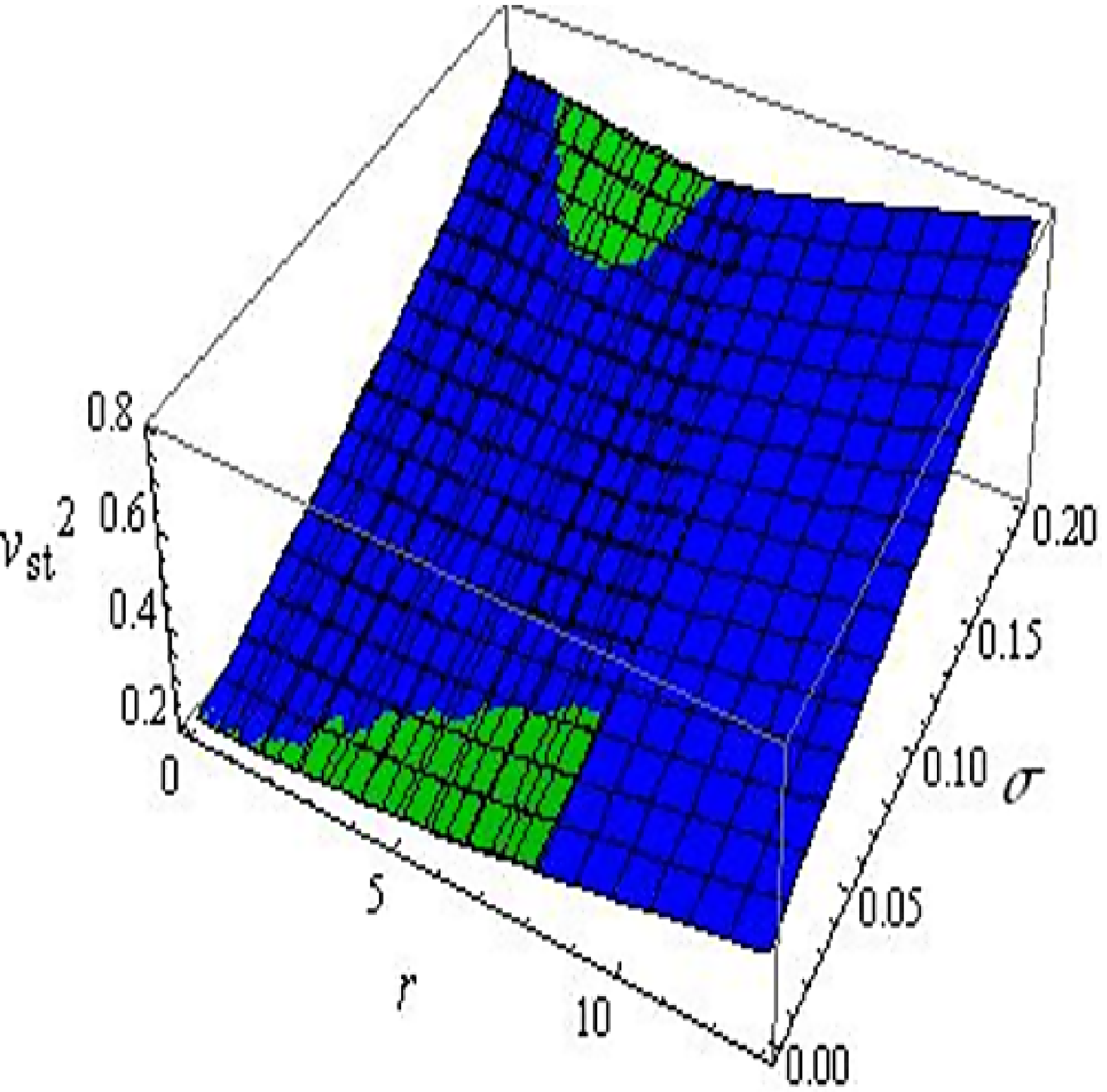,
width=0.45\linewidth}\\ \epsfig{file=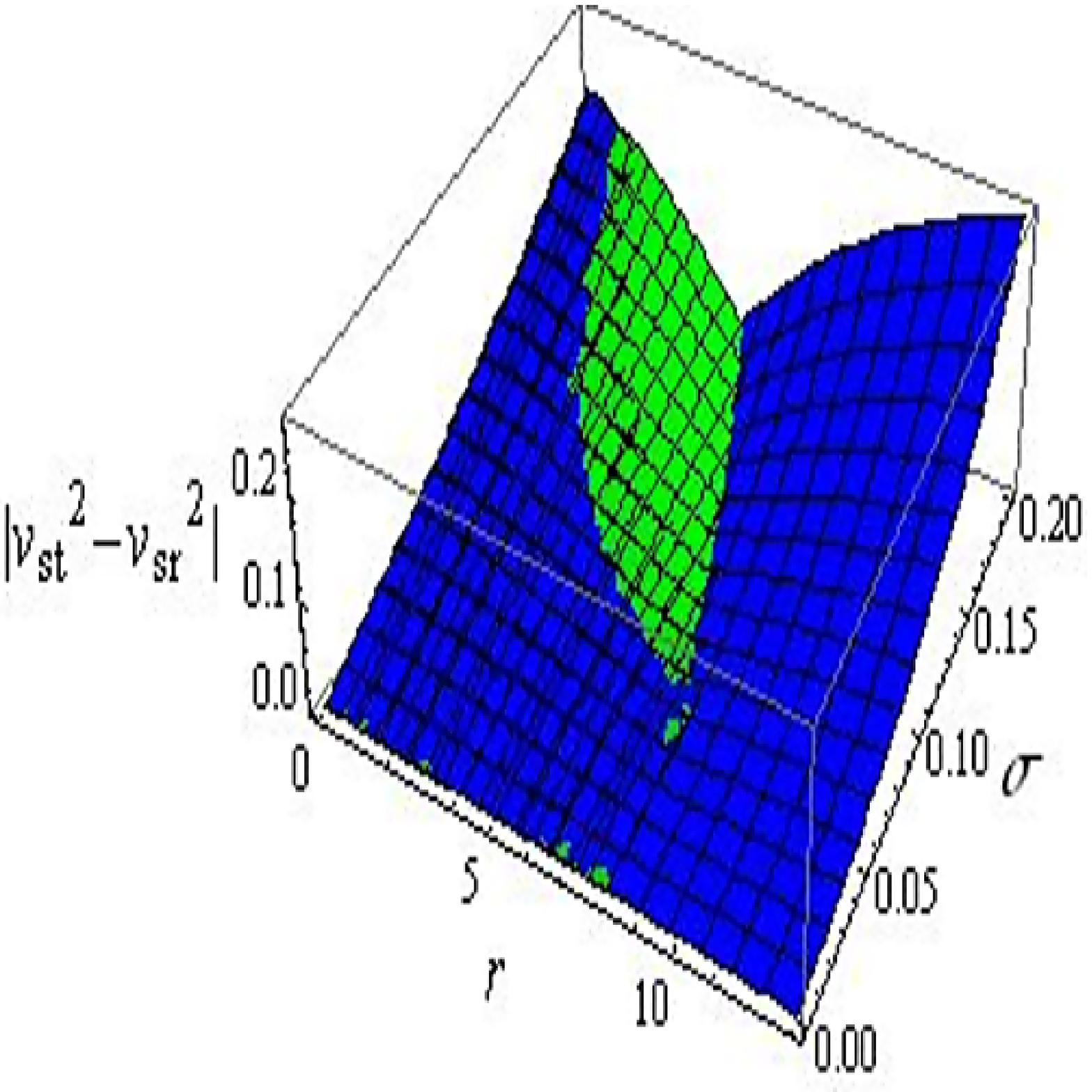, width=0.45\linewidth}
\caption{Plots of $v_{sr}^{2}$, $v_{st}^{2}$ and
$|v_{st}^{2}-v_{sr}^{2}|$ for Her X-I (green) and PSR J 1416-2230
(blue) corresponding to solution $\mathbf{II}$.}
\end{figure}
\begin{figure}\center \epsfig{file=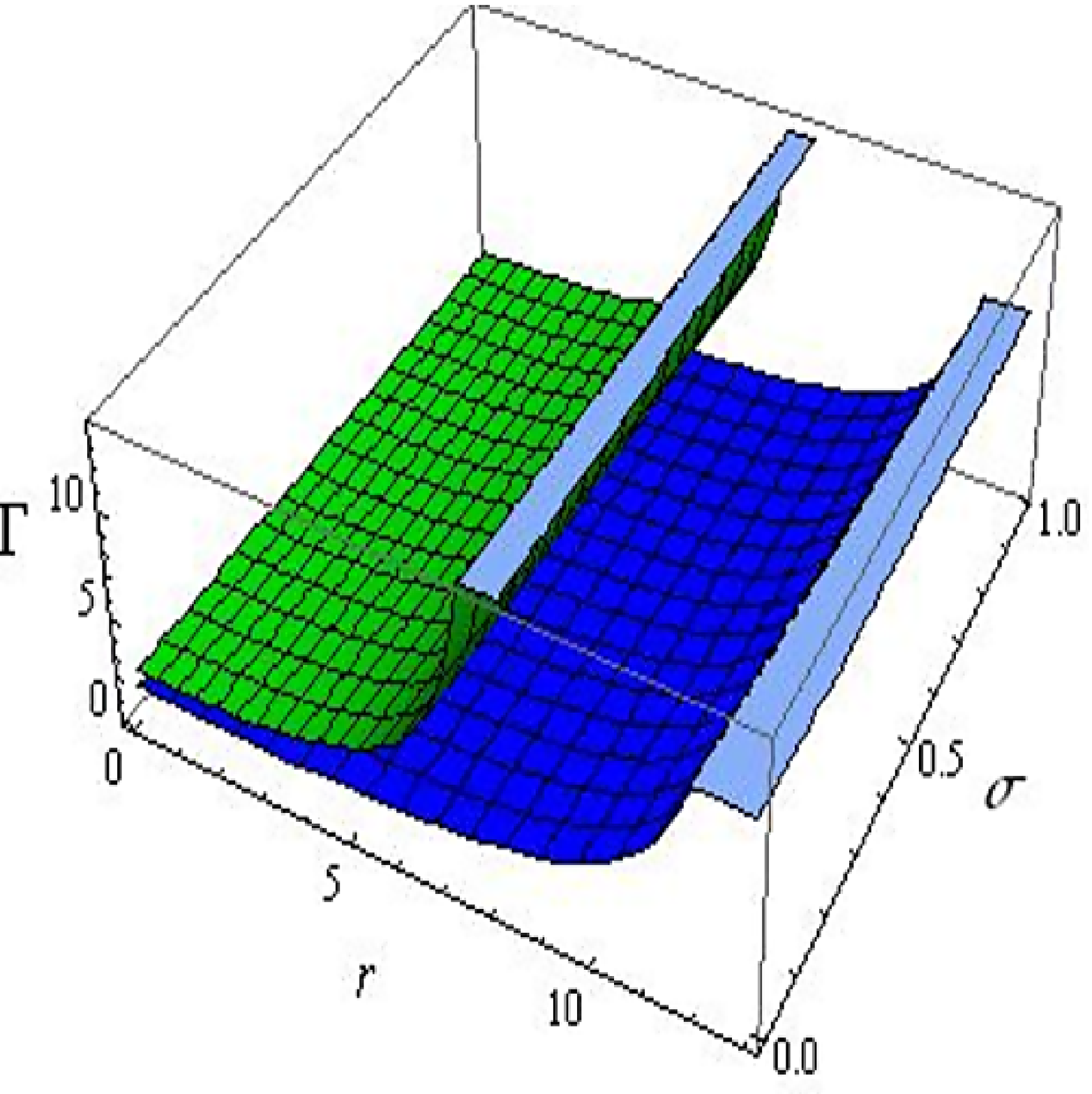,width=0.45\linewidth}
\epsfig{file=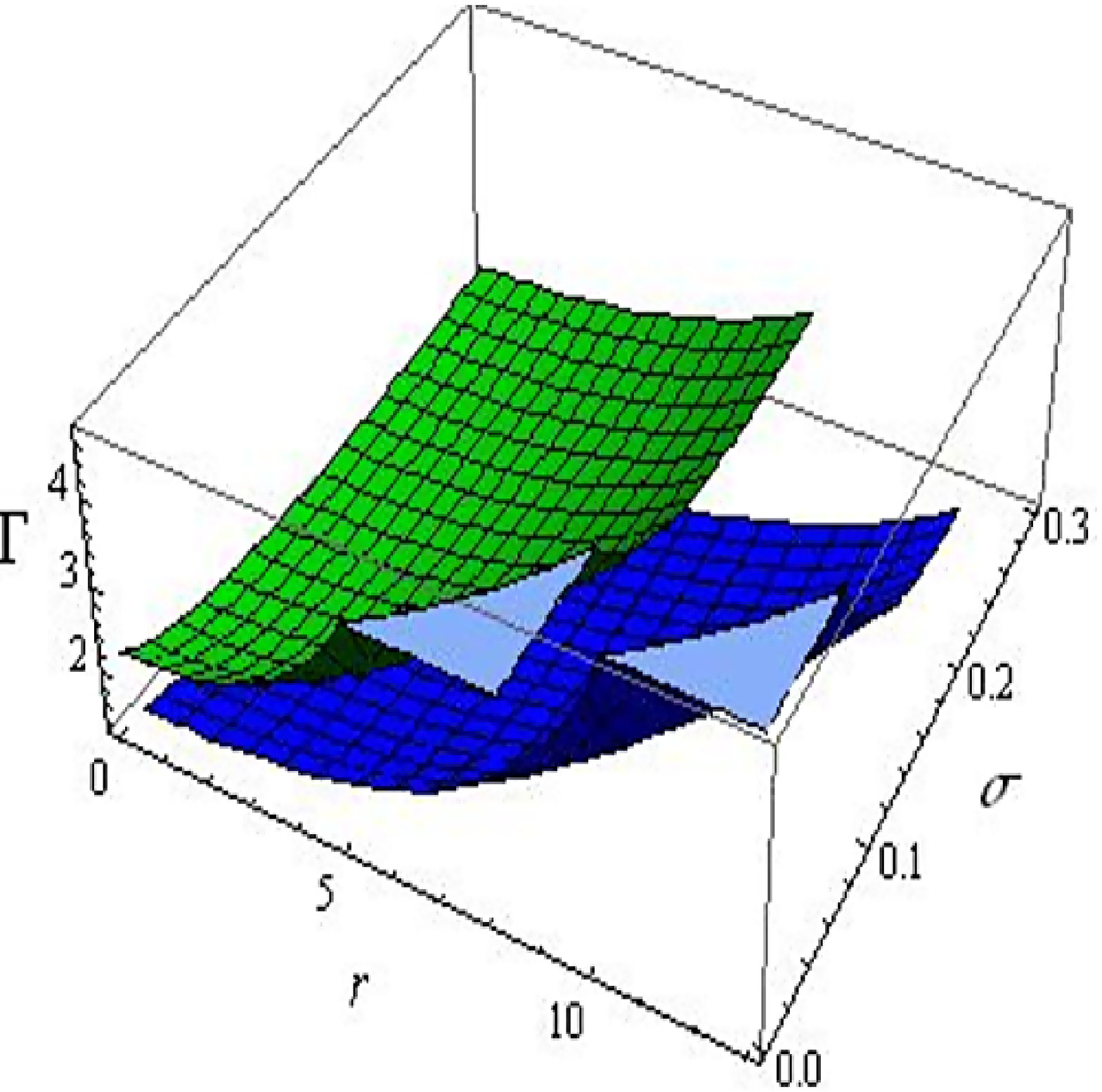,width=0.45\linewidth}
\\ \caption{Plots of
adiabatic index for Her X-I (green) and PSR J 1416-2230 (blue)
corresponding to solutions $\mathbf{I}$ (left) and $\mathbf{II}$
(right).}
\end{figure}

\subsection{Adiabatic Index}

The adiabatic index $\Gamma$ as a stiffness parameter has
significant importance to study the stable behavior of relativistic
stellar structure. Chandrasekhar \cite{25} and many researchers
\cite{26}-\cite{28} discussed the stability of gaseous stars against
radial adiabatic perturbation. It is found that $\Gamma$ should be
greater than $\frac{4}{3}$ in the interior of stable isotropic
celestial objects. For anisotropic sphere, the adiabatic index is
defined by
\begin{equation}
\Gamma_{r}=\frac{\hat{\rho}+\hat{p}_{r}}{\hat{p}_{r}}v_{sr  }^{2}.
\end{equation}
The graphical representation of $\Gamma$ is displayed in Figure
\textbf{11} which shows that our constructed anisotropic models
indicate dynamical stable behavior as the value of
$\Gamma>\frac{4}{3}$ throughout the domain. Moreover, it is noted
that the adiabatic index becomes undefined at the boundary due to
the vanishing of radial pressure at $r=R$.

\section{Concluding Remarks}

The formulation of analytic solutions describing the interior
constituents of self-gravitating objects has captured the attention
of many researchers. In this context, gravitational decoupling by
MGD technique has effectively been used to explore anisotropic
solutions for matter sources. Here, matter sources interact
gravitationally with no exchange of energy. To resolve this issue,
an extended version of the MGD was presented \cite{14} which enables
the transformation of energy between matter contents. In this paper,
we have derived exact spherical anisotropic solutions from a known
isotropic model through the EGD technique. To include the effects of
anisotropy, we have deformed both temporal as well as radial metric
potentials and successfully decoupled the field equations. The
Bianchi identities for both non-generic matter sources have been
studied which provide a particular relation between them as
displayed in Eqs.(\ref{26}) and (\ref{27}). Moreover, the unknown
constants of the models are computed through the matching of
interior anisotropic solutions with the Schwarzschild spacetime.

To examine the consistency of the EGD approach, we have first
considered the isotropic Tolman IV solution and incorporated the
effects of anisotropy by adding an additional matter source in the
perfect fluid configuration. In order to evaluate the exact
anisotropic solutions, we have used a barotropic equation of state
for $\vartheta_{\alpha}^{\beta}$ and imposed the conditions on
pressure and energy density which yield the solutions \textbf{I} and
\textbf{II}, respectively. We have investigated physical properties
of the constructed models through the graphical portray of matter
variables, anisotropic factor and energy conditions. It is observed
that both solutions are physically acceptable as they satisfy all
the essential viability conditions for the star candidates ``Her
X-I" and ``PSR J 1416-2230". Moreover, the stability criteria
provided by the speed of sound constraint and adiabatic index are
fulfilled for proposed models that depict the potentially stable
structure of compact stars. It is worthwhile to mention here that
the both derived anisotropic solutions given in
Eqs.(\ref{49})-(\ref{52}) and (\ref{62})-(\ref{63}) satisfy the
field equations (\ref{9})-(\ref{11}) as shown in Appendix A.

Ovalle et al. \cite{17a} derived anisotropic spherical solutions
from perfect Tolman IV model using the MGD approach but the energy
bounds as well as stability conditions were not investigated for
their solutions. Sharif and Sadiq \cite{16} extended Krori-Barua
solution for charged spherical system to anisotropic domain and
deduced that the first solution corresponding to pressure constraint
exhibits the stable behavior whereas the second solution violates
the viability criteria. In a recent paper, Sharif and Saba
\cite{16a} evaluated anisotropic spherical solutions using MGD in
the background of $f(\mathcal{G})$ gravity and found only one viable
as well as stable solution. It is worth mentioning here that our
both anisotropic models indicate the consistent behavior and satisfy
the stability criteria. We would like to re-iterate that EGD
technique is more powerful as this can provide transformation of
energy between matter contents and helps to study the physical
characteristics of self-gravitating objects.

\section*{Appendix A}
\renewcommand{\theequation}{A\arabic{equation}}
\setcounter{equation}{0}

Here we show that the anisotropic solutions given in
Eqs.(\ref{49})-(\ref{52}) and (\ref{62})-(\ref{63}) satisfy the
field equations (\ref{9})-(\ref{11}). Moreover, the conservation
equations (\ref{23}) and (\ref{28}) are verified. For this purpose,
we proceed as follows. Differentiating Eqs.(\ref{51}) and (\ref{52})
with respect to $r$, it follows that
\begin{eqnarray}\nonumber
\eta^{'}&=&\frac{2 r}{\mathcal{A}^2
\left(\frac{r^2}{\mathcal{A}^2}+1\right)}+\sigma\frac{1}{\varepsilon}\left(2r
\left(\mathcal{A}^6+8\mathcal{C}^2r^4+6
\mathcal{A}^2r^2\left(\mathcal{C}^2+2r^2\right)+\mathcal{A}^4
\left(2\mathcal{C}^2\right.\right.\right.\\
\label{A2}&+&\left.\left.\left.7r^2\right)\right)\sqrt{2}\mathcal{A}
\left(\mathcal{A}^2+2\mathcal{C}^2\right)\left(\mathcal{A}^4
+5\mathcal{A}^2r^2+6r^4\right)\tan^{-1}\left(\frac{\sqrt{2}
r}{\mathcal{A}}\right)\right), \\ \nonumber \eta^{''}&=&-((2 \sigma
\left(\mathcal{A}^2+r^2\right) \bigg(\sqrt{2} \mathcal{A}
\left(\mathcal{A}^2+2 \mathcal{C}^2\right) \left(5 \mathcal{A}^2+12
r^2\right)\tan ^{-1}\left(\frac{\sqrt{2} r}{\mathcal{A}}\right)\\
\nonumber&-&2 r \left(3 \mathcal{A}^2+2 \mathcal{C}^2\right) \left(3
\mathcal{A}^2+10 r^2\right)\bigg))(\sigma \left(\mathcal{A}^2+2
r^2\right) (-2 r \left(\mathcal{A}^2+2
\mathcal{C}^2\right)\\\nonumber&+&\sqrt{2} \mathcal{A}
\left(\mathcal{A}^2+2 \mathcal{C}^2\right) \tan
^{-1}\left(\frac{\sqrt{2} r}{\mathcal{A}}\right)+4 r^3)+8 r
\left(\mathcal{A}^2+r^2\right) (\mathcal{C}^{2}-r^{2}))^{-1}\\
\nonumber&-&2 \left(\mathcal{A}^2+r^2\right)+(2 \sigma (2 r
(\mathcal{A}^6+\mathcal{A}^4 \left(2 \mathcal{C}^2+7
r^2\right)+6 \mathcal{A}^2 r^2 \left(\mathcal{C}^2+2 r^2\right)\\
\nonumber&+&8 \mathcal{C}^2 r^4)-\sqrt{2} \mathcal{A}
\left(\mathcal{A}^2+2 \mathcal{C}^2\right) \left(\mathcal{A}^4+5
\mathcal{A}^2 r^2+6 r^4\right) \tan ^{-1}\left(\frac{\sqrt{2}
r}{\mathcal{A}}\right)))\\ \nonumber&\times&(\sigma
\left(\mathcal{A}^2+2 r^2\right) (-2 r \left(\mathcal{A}^2+2
\mathcal{C}^2\right)+\sqrt{2} \mathcal{A} \left(\mathcal{A}^2+2
\mathcal{C}^2\right) \tan ^{-1}\left(\frac{\sqrt{2}
r}{\mathcal{A}}\right)\\ \nonumber&+&4 r^3)+8 r
\left(\mathcal{A}^2+r^2\right) (\mathcal{C}^{2}-r^{2}))^{-1}-(4
\sigma \left(\mathcal{A}^2+r^2\right)(\sqrt{2} \mathcal{A}
(\mathcal{A}^2\\ \nonumber&+&2 \mathcal{C}^2) \left(\mathcal{A}^4+5
\mathcal{A}^2 r^2+6 r^4\right) \tan ^{-1}\left(\frac{\sqrt{2}
r}{\mathcal{A}}\right)-2 r (\mathcal{A}^6+\mathcal{A}^4 (2
\mathcal{C}^2\\ \nonumber&+&7 r^2)+6 \mathcal{A}^2 r^2
\left(\mathcal{C}^2+2 r^2\right)+8 \mathcal{C}^2 r^4)) (2
(\mathcal{A}^2 \left(\mathcal{C}^2-3 r^2\right)+(\sigma -1)\\
\nonumber&\times& r^2 \left(5 r^2-3 \mathcal{C}^2\right))+\sqrt{2}
\mathcal{A} \sigma  r \left(\mathcal{A}^2+2 \mathcal{C}^2\right)
\tan ^{-1}\left(\frac{\sqrt{2} r}{\mathcal{A}}\right)))(r \\
\nonumber&\times&(\sigma \left(\mathcal{A}^2+2 r^2\right) (-2 r
\left(\mathcal{A}^2+2 \mathcal{C}^2\right)+\sqrt{2}\mathcal{A}
\left(\mathcal{A}^2+2 \mathcal{C}^2\right) \tan ^{-1}(\frac{\sqrt{2}
r}{\mathcal{A}})\\ \nonumber&+&4 r^3)+8 r
\left(\mathcal{A}^2+r^2\right) (\mathcal{C}^{2}-r^{2})
)^2)^{-1}+(\sigma \left(\mathcal{A}^2+r^2\right)(2 r
(\mathcal{A}^6\\ \nonumber&+&\mathcal{A}^4 \left(2 \mathcal{C}^2+7
r^2\right)+6 \mathcal{A}^2 r^2 \left(\mathcal{C}^2+2 r^2\right)+8
\mathcal{C}^2 r^4)-\sqrt{2} \mathcal{A} \left(\mathcal{A}^2+2
\mathcal{C}^2\right)\\ \nonumber&\times& \left(\mathcal{A}^4+5
\mathcal{A}^2 r^2+6 r^4\right) \tan ^{-1}(\frac{\sqrt{2}
r}{\mathcal{A}})))(r^2 (\sigma \left(\mathcal{A}^2+2 r^2\right) (-2
r \\ \nonumber&\times&\left(\mathcal{A}^2+2
\mathcal{C}^2\right)+\sqrt{2} \mathcal{A} \left(\mathcal{A}^2+2
\mathcal{C}^2\right) \tan ^{-1}\left(\frac{\sqrt{2}
r}{\mathcal{A}}\right)+4 r^3)+8 r\\ \label{A3}&+&
\left(\mathcal{A}^2+r^2\right)
(\mathcal{C}^{2}-r^{2})))^{-1}+4r^2)(\left(\mathcal{A}^2+r^2\right)^2)^{-1},\\
\nonumber \chi^{'}e^{-\chi}&=&(\sqrt{2} \mathcal{A} \sigma
\left(\mathcal{A}^2+2 \mathcal{C}^2\right) \left(\mathcal{A}^2+2
r^2\right)^2 \tan ^{-1}\left(\frac{\sqrt{2} r}{\mathcal{A}}\right)-2
r \sigma \left(\mathcal{A}^2+2 r^2\right)\\ \nonumber&\times&
\left(\mathcal{A}^4+2 \mathcal{A}^2 \left(\mathcal{C}^2+2
r^2\right)+8 r^4\right)+16 r^3 \left(\mathcal{A}^4+\mathcal{A}^2
\left(\mathcal{C}^2+2 r^2\right)+2 r^4\right))\\
\label{A1}&\times&(8 \mathcal{C}^2 r^2 \left(\mathcal{A}^2+2
r^2\right)^2)^{-1}.
\end{eqnarray}
The derivatives of Eqs.(\ref{39}), (\ref{42}) and (\ref{50}) lead to
\begin{eqnarray}\label{A4}
\xi^{'}&=&\frac{2 r}{a^2 \left(\frac{r^2}{a^2}+1\right)},\\
\label{A5} p^{'}&=&-\frac{2 r \left(a^2+2 c^2\right)}{c^2
\left(a^2+2 r^2\right)^2}=-\vartheta_{1}^{'1},\\
\nonumber h^{'}&=&\frac{1}{\varepsilon}\left(2 r
\left(\mathcal{A}^6+8\mathcal{C}^2r^4+6
\mathcal{A}^2r^2\left(\mathcal{C}^2+2r^2\right)+\mathcal{A}^4
\left(2\mathcal{C}^2+7r^2\right)\right)-\sqrt{2}\mathcal{A} \right.\\
\label{A6}&\times&\left.\left(\mathcal{A}^2+2\mathcal{C}^2\right)
\left(\mathcal{A}^4+5\mathcal{A}^2
r^2+6r^4\right)\tan^{-1}\left(\frac{\sqrt{2}
r}{\mathcal{A}}\right)\right),
\end{eqnarray}
The field equation (\ref{9}) can be written as
\begin{eqnarray}\label{A7}
\hat{\rho}&=&\frac{1}{r^2}+e^{-\chi}
\left(\frac{\chi^{'}}{r}-\frac{1}{r^2}\right).
\end{eqnarray}
Using Eq.(\ref{A1}), the right side of the above equation takes the
form
\begin{eqnarray*}
\frac{3\mathcal{A}^4+\mathcal{A}^2(3\mathcal{C}^2+7r^2)
+2r^2(\mathcal{C}^2+3r^2)}{\mathcal{C}^2(\mathcal{A}^2+2r^2)^2}
+\sigma\left(\frac{\mathcal{C}^2-\mathcal{A}^2-3r^2}
{\mathcal{C}^2(\mathcal{A}^2+2r^2)}\right),
\end{eqnarray*}
which is just equal to $\hat{\rho}$ given in Eq.(\ref{56}). Consider
the right side of Eq.(\ref{10}) as
\begin{eqnarray*}
-\frac{1}{r^2}+e^{-\chi}
\left(\frac{\eta^{'}}{r}+\frac{1}{r^2}\right).
\end{eqnarray*}
Using Eqs.(\ref{A2}) and (\ref{A1}), we have
\begin{eqnarray*}
\frac{\mathcal{C}^2-\mathcal{A}^2-3r^2}
{\mathcal{C}^2(\mathcal{A}^2+2r^2)}(1 - \sigma),
\end{eqnarray*}
which is equal to $\hat{p}_{r}=p+\sigma\vartheta_{1}^{1}$ given in
Eq.(\ref{57}). Similarly, the right side of Eq.(\ref{11}) is
expressed as
\begin{eqnarray*}
\frac{e^{-\chi}}{4} \left(2\eta^{''}+\eta^{'2}-\eta^{'}\chi^{'}
+\frac{2}{r}(\eta^{'}-\chi^{'})\right),
\end{eqnarray*}
which, through Eqs.(\ref{A2}) and (\ref{A1}) gives rise to
\begin{eqnarray*}\nonumber
&&\frac{\mathcal{C}^2-\mathcal{A}^2-3r^2}
{\mathcal{C}^2(\mathcal{A}^2+2r^2)}+\sigma
(2r(16r^4(\mathcal{C}^4-3\mathcal{C}^2r^2
(-1+\sigma)+3r^4(-1+\sigma))\\
\nonumber&+&\mathcal{A}^8(1-2\sigma)+4\mathcal{A}^4
(\mathcal{C}^2r^2(21-8\sigma)+\mathcal{C}^4(-1+\sigma)
-r^4(11+\sigma))\\
\nonumber&+&4
\mathcal{A}^2r^2(2\mathcal{C}^2r^2(17-10\sigma)+12r^4(-2+\sigma
)+\mathcal{C}^4(1+2\sigma))\\
\nonumber&-&\mathcal{A}^6(2\mathcal{C}^2
(-6+\sigma)+r^2(1+12\sigma)))+\sqrt{2} \mathcal{A}
(\mathcal{A}^2+2\mathcal{C}^2)(\mathcal{A}^2+2 r^2)\\
\nonumber&\times&(-(\mathcal{A}^2+2\mathcal{C}^2)
(\mathcal{A}^2+3r^2)+2(\mathcal{A}^4+6r^4-\mathcal{A}^2
(\mathcal{C}^2-6r^2))\sigma)\tan^{-1}(\frac{\sqrt{2}
r}{\mathcal{A}}))\\ \nonumber&\times&\left(2
\mathcal{C}^2\left(\mathcal{A}^2+2r^2\right)^2[8(\mathcal{C}-r)r
(\mathcal{C}+r)
\left(\mathcal{A}^2+r^2\right)+\left(\mathcal{A}^2+2r^2\right)
\right.\\ &\times&\left.\sigma
(-2\left(\mathcal{A}^2+2\mathcal{C}^2\right)r+4
r^3+\sqrt{2}\mathcal{A}\left(\mathcal{A}^2+2\mathcal{C}^2\right)
\tan^{-1}\left(\frac{\sqrt{2} r}{\mathcal{A}}\right))]\right)^{-1}.
\end{eqnarray*}
This is exactly the same as $\hat{p}_{t}=p+\sigma\vartheta_{2}^{2}$
given in Eq.(\ref{59}). Hence, the solution $\textbf{I}$ satisfies
the field equations.  Using Eqs.(\ref{18}) and (\ref{19}), we obtain
the following constraint
\begin{eqnarray}\label{A5}
-\frac{1}{r^2}+e^{-\mu}
\left(\frac{\xi^{'}}{r}+\frac{1}{r^2}\right)&=&e^{-\mu}\left(\frac{\xi^{''}}{2}+\frac{\xi^{'2}}{4}
-\frac{\xi^{'}\mu^{'}}{4}+\frac{1}{2r}(\xi^{'}-\mu^{'})\right),
\end{eqnarray}
where its left side reads
\begin{eqnarray}
\frac{1}{2} \left(-\frac{\mathcal{A}^2+2
\mathcal{C}^2}{\mathcal{C}^2 \left(\mathcal{A}^2+2
r^2\right)}+\frac{3}{\mathcal{C}^2}-\frac{4}{r^2}\right),
\end{eqnarray}
which turns out to be the same to the right side of Eq.(\ref{A5})
after inserting the values. This indicates that solutions of the
field equations satisfy the constraints derived from them. The
conservation equation (\ref{23}) can be written as
\begin{equation}\label{A8}
p^{'}+\frac{\xi^{'}}{2}(\rho+p)+\frac{\sigma
h^{'}}{2}(\rho+p)=-\sigma
\left(\vartheta_{1}^{1'}+\frac{\eta^{'}}{2}(\vartheta_{1}^{1}
-\vartheta_{0}^{0})+\frac{2}{r}(\vartheta_{1}^{1}-\vartheta_{2}^{2})\right).
\end{equation}
Using Eqs.(\ref{A4})-(\ref{A6}) and (\ref{41})-(\ref{42}), its left
side is evaluated as
\begin{eqnarray*}
&-&(\sigma  \left(\mathcal{A}^2+2 \mathcal{C}^2\right) (\sqrt{2}
\mathcal{A} \left(\mathcal{A}^2+2 \mathcal{C}^2\right)
\left(\mathcal{A}^4+5 \mathcal{A}^2 r^2+6 r^4\right) \tan
^{-1}\left(\frac{\sqrt{2} r}{\mathcal{A}}\right)-2 r
\\ \nonumber&\times&\left(\mathcal{A}^6+\mathcal{A}^4 \left(2 \mathcal{C}2+7
r^2\right)+6 \mathcal{A}^2 r^2 \left(\mathcal{C}^2+2 r^2\right)+8
\mathcal{C}^2 r^4\right)))(\mathcal{C}^2 r \left(\mathcal{A}^2+2
r^2\right)^2 \\ \nonumber&\times&(\sigma \left(\mathcal{A}^2+2
r^2\right) \left(-2 r \left(\mathcal{A}^2+2
\mathcal{C}^2\right)+\sqrt{2} \mathcal{A} \left(\mathcal{A}^2+2
\mathcal{C}^2\right) \tan ^{-1}\left(\frac{\sqrt{2}
r}{\mathcal{A}}\right)+4 r^3\right)\\ \nonumber&+&8 r
\left(a^2+r^2\right) (\mathcal{C}-r) (\mathcal{C}+r)))^{-1},
\end{eqnarray*}
which is equal to the right side of Eq.(\ref{A8}). Similarly, the
second conservation equation can be satisfied. Hence, the
anisotropic solution $\textbf{I}$ satisfies the conservation
equations.

Now, we repeat the same procedure for the second anisotropic
solution. The metric potentials take the form
\begin{eqnarray}\label{A9}
e^{-\chi}&=&\frac{(1-\frac{r^2}{\mathcal{C}^2})(1+\frac{r^2}
{\mathcal{A}^2})}{1+\frac{2r^2}{\mathcal{A}^2}}+\sigma
(\frac{r^2\left(\mathcal{A}^2+\mathcal{C}^2+r^2\right)}
{\mathcal{C}^2\left(\mathcal{A}^2+2 r^2\right)}),\\
\nonumber \eta&=&\sigma\bigg(\frac{1}{2}(-\frac{2
\ln\left(\mathcal{A}^2+r^2\right)}{\sigma }-\frac{2
\ln\left(\mathcal{A}^2+2 r^2\right)}{-1+\sigma
}+\left(\left(\mathcal{A}^2(-1+\sigma )+\mathcal{C}^2 (-1+3\sigma)
\right.\right.\\\nonumber&+&\left.\left.(-1+2 \sigma )\varrho\right)
\ln\left(\mathcal{C}^2 (1+\sigma )-\varrho+\left(\mathcal{A}^2+2
r^2\right)(-1+\sigma)\right)\right)\left(\varrho\sigma(-1+\sigma)
\right)^{-1}\\ \nonumber&+&\left(\left(\varrho(-1+2\sigma)
+\mathcal{C}^2 (1-3 \sigma )-\mathcal{A}^2 (-1+\sigma)\right)
\ln\left(\left(\mathcal{A}^2+2 r^2\right) (-1+\sigma )\right.\right.\\
\label{A10}&+&\left.\left.\mathcal{C}^2
(1+\sigma)+\varrho\right)\right)\left((-1+\sigma)\sigma
\varrho\right)^{-1})\bigg)+\ln\left(\mathcal{B}^2(1+\frac{r^2}{\mathcal{A}^2})
\right),
\end{eqnarray}
The derivatives of the above equations lead to
\begin{eqnarray} \label{A11}
\chi^{'}e^{-\chi}&=&-(2 r (\sigma -1)
\left(\mathcal{A}^4+\mathcal{A}^2 \left(\mathcal{C}^2+2 r^2\right)+2
r^4\right))(\mathcal{C}^2 \left(\mathcal{A}^2+2 r^2\right)^2)^{-1},\\
\nonumber \eta^{'}&=&(2 r \left(\left(\mathcal{A}^2+2 r^2\right)
(\mathcal{C}^{2}-r^{2})+\sigma \left(\mathcal{A}^4+\mathcal{A}^2
\left(\mathcal{C}^2+2 r^2\right)+2 r^4\right)\right))\\
\nonumber&\times&(\left(\mathcal{A}^2+2 r^2\right) \left(r^2 \sigma
\left(\mathcal{A}^2+\mathcal{C}^2+r^2\right)+\left(\mathcal{A}^2+r^2\right)
(\mathcal{C}-r) (\mathcal{C}+r)\right))^{-1},\\
\label{A12}\\\nonumber \eta^{''}&=& (2 \mathcal{A}^8 \sigma
\left(\mathcal{C}^2-r^2 (\sigma -1)\right)+2 \mathcal{A}^6
(\mathcal{C}^4 (\sigma +1)+\mathcal{C}^2 r^2 ((3-2 \sigma ) \sigma
-2)\\ \nonumber&+&r^4 ((6-7 \sigma ) \sigma +1))-2 \mathcal{A}^4 r^2
(\mathcal{C}^4
(\sigma (\sigma +4)-3)+\mathcal{C}^2 r^2 (\sigma (13 \sigma -11)\\
\nonumber&+&6)+r^4 (\sigma -1) (10 \sigma +3))-4 \mathcal{A}^2 r^4
\sigma(\mathcal{C}^4 (3 \sigma +5)+2 \mathcal{C}^2 r^2 (2 \sigma
-1)\\\nonumber&+&3 r^4 (\sigma -1))+8 r^6 \left(\mathcal{C}^4
(-(\sigma +1))+\mathcal{C}^2 r^2 (\sigma -2) (\sigma -1)-r^4 (\sigma
-1)^2\right))\\
\nonumber&\times&(\left(\mathcal{A}^2+2 r^2\right)^2(r^2 \sigma
\left(\mathcal{A}^2+\mathcal{C}^2+r^2\right)+\left(\mathcal{A}^2+r^2\right)
(\mathcal{C}-r)(\mathcal{C}+r))^2)^{-1}.\\ \label{A13}
\end{eqnarray}
Using Eq.(\ref{A11}), the right side of Eq.(\ref{A7}) is evaluated
as
\begin{eqnarray*}
\frac{3\mathcal{A}^4+\mathcal{A}^2(3\mathcal{C}^2+7r^2)
+2r^2(\mathcal{C}^2+3r^2)}{\mathcal{C}^2(\mathcal{A}^2+2r^2)^2}(1-\sigma),
\end{eqnarray*}
which is similar to $\hat{\rho}$ given in (\ref{67}). Through
Eqs.(\ref{A11}) and (\ref{A12}), the right side of Eq.(\ref{10})
reads
\begin{eqnarray*}
\frac{\mathcal{C}^2-\mathcal{A}^2-3r^2}
{\mathcal{C}^2(\mathcal{A}^2+2r^2)}+\sigma(\frac{3\mathcal{A}^4
+\mathcal{A}^2(3\mathcal{C}^2+7r^2)
+2r^2(\mathcal{C}^2+3r^2)}{\mathcal{C}^2(\mathcal{A}^2+2r^2)^2}),
\end{eqnarray*}
which is similar to $\hat{p}_{r}$ given in (\ref{68}). Similarly,
the right side of Eq.(\ref{11}) takes the form
\begin{eqnarray*}
&&\frac{\mathcal{C}^2-\mathcal{A}^2-3r^2}{\mathcal{C}^2
(\mathcal{A}^2+2r^2)}+\sigma(3\mathcal{A}^6+3\mathcal{A}^4\mathcal{C}^2
+8\mathcal{A}^4r^2-2\mathcal{A}^2\mathcal{C}^2r^2+18\mathcal{A}^2r^4+12
r^6\\\nonumber&+&\frac{\left(\mathcal{A}^2+2\mathcal{C}^2\right)r^2
\left(\mathcal{A}^6+\mathcal{A}^2
r^4-2\mathcal{C}^2r^4+\mathcal{A}^4
\left(\mathcal{C}^2+2r^2\right)\right)}{(\mathcal{C}-r)(\mathcal{C}+r)
\left(\mathcal{A}^2+r^2\right)+r^2
\left(\mathcal{A}^2+\mathcal{C}^2+r^2\right)\sigma})(\mathcal{C}^2
\left(\mathcal{A}^2+2 r^2\right)^3)^{-1},
\end{eqnarray*}
which is similar to $\hat{p}_{t}$ given in (\ref{69}). The left side
of conservation equation (\ref{A8}) is computed as
\begin{eqnarray*}
\frac{2 r \sigma  \left(\mathcal{A}^2+2 \mathcal{C}^2\right)
\left(\mathcal{A}^6+\mathcal{A}^4 \left(\mathcal{C}^2+2
r^2\right)+\mathcal{A}^2 r^4-2 \mathcal{C}^2
r^4\right)}{\mathcal{C}^2 \left(\mathcal{A}^2+2 r^2\right)^3
\left(r^2 \sigma
\left(\mathcal{A}^2+\mathcal{C}^2+r^2\right)+\left(\mathcal{A}^2+r^2\right)
(\mathcal{C}^{2}-r^{2})\right)},
\end{eqnarray*}
which is equal to its right side. Similarly, the second conservation
equation is satisfied. Hence, both anisotropic solutions satisfy the
field equations as well as conservation equations.

\vspace{5cm}

\textbf{Acknowledgement}

\vspace{0.5cm}

One of us (QM) would like to thank the Higher Education Commission,
Islamabad, Pakistan for its financial support through the
\emph{Indigenous Ph.D. Fellowship, Phase-II, Batch-III}.


\vspace{0.5cm}

\begin{thebibliography}{40}

\bibitem{1} Schwarzschild, K.: Kl. Math. Phys.
\textbf{24}(1916)424.

\bibitem{2} Tolman, R.C.: Phys. Rev. \textbf{55}(1939)364.

\bibitem{3} Lemaitre, G.: Ann. Soc. Sci. Bruxells
\textbf{A53}(1933)51.

\bibitem{4} Ruderman, R.: Ann. Rev. Astron. Astrophys. \textbf{10}(1972)427.

\bibitem{4'} Bowers, R.L. and Liang, E.P.T.: Astrophys. J.
\textbf{188}(1974)657.

\bibitem{4a} Abbas, G. et al.: Astrophys. Space Sci. \textbf{357}(2015)158.

\bibitem{4b} Tripathy, S.K. and Mishra, B.: Eur. Phys. J. Plus
\textbf{131}(2016)273.

\bibitem{4''} Murad, M.H.: Astrophys. Space Sci.
\textbf{20}(2016)361.

\bibitem{4'''} Maurya, S.K. and Maharaj, S.D.: Eur. Phys. J. C
\textbf{77}(2017)328.

\bibitem{4''''} Matondo, D.K., Maharaj, S.D. and Ray, S.: Eur. Phys. J. C
\textbf{78}(2018)437.

\bibitem{9} Ovalle, J.: Mod. Phys. Lett. A \textbf{23}(2008)3247.

\bibitem{10} Ovalle, J. and Linares, F.: Phys. Rev. D
\textbf{88}(2013)104026.

\bibitem{17} Contreras, E. and Bargue$\tilde{n}$o, P.: Eur. Phys. J. C
\textbf{78}(2018)558.

\bibitem{20} Ovalle, J.: Phys. Rev. D \textbf{95}(2017)104019.

\bibitem{17a} Ovalle, J. et al.: Eur. Phys. J. C
\textbf{78}(2018)122.

\bibitem{16} Sharif, M. and Sadiq, S.: Eur. Phys. J. C
\textbf{78}(2018)410.

\bibitem{16c} Gabbanelli, L., Rincon, A. and Rubio, C.:
Eur. Phys. J. C \textbf{78}(2018)370.

\bibitem{16d} Graterol, R.P.: Eur. Phys. J. Plus \textbf{133}(2018)244.

\bibitem{16a} Sharif, M. and Saba, S.: Eur. Phys. J. C
\textbf{78}(2018)921.

\bibitem{16b} Sharif, M. and Waseem, A.: Ann. Phys. \textbf{405}(2019)14.

\bibitem{17'} Sharif, M. and Ama-Tul-Mughani, Q.: Int. J. Geom. Methods Mod. Phys.
\textbf{16}(2019)1950187; Mod. Phys. Lett. A (to appear, 2020).

\bibitem{17''} Casadio, R. et al.: Eur. Phys. J. C
\textbf{79}(2019)826.

\bibitem{11} Casadio, R., Ovalle, J. and da Rocha, R.: Class. Quantum Grav.
\textbf{32}(2015)215020.

\bibitem{14} Ovalle, J.: Phys. Lett. B \textbf{788}(2019)213.

\bibitem{14'} Contreras, E. and Bargue$\tilde{n}$o, P.: Class. Quantum
Grav. \textbf{36}(2019)215009.

\bibitem{21} Ovalle, J. et al.: Eur. Phys. J. C \textbf{78}(2018)960.

\bibitem{23a} Tolman, R. C.: Phys. Rev. \textbf{55}(1939)364.

\bibitem{23b} Oppenheimer, J.R. and Volkoff. G.M.: Phys. Rev. \textbf{55}(1939)374.

\bibitem{24a} Abreu, H., Hernandez, H. and Nunez, L.A.: Class. Quantum Gravit.
\textbf{24}(2007)4631.

\bibitem{24b} Herrera, L.: Phys. Lett. A \textbf{165}(1992)206.

\bibitem{24c} Andreasson, H.: Commun. Math. Phys.
\textbf{288}(2009)715.

\bibitem{25} Chandrasekhar, S.: Astrophys. J. \textbf{140}(1964)417.

\bibitem{26} Heintzmann, H.: Hillebrandt, W.: Astron. Astrophys.
\textbf{38}(1975)51.

\bibitem{27} Hillebrandt, W. and Steinmetz, K.O.: Astron. Astrophys.
\textbf{53}(1976)283.

\bibitem{28} Bombaci, I.: Astron. Astrophys. \textbf{305}(1996)871.

\end{thebibliography}
\end{document}